\documentclass[acmsmall,screen]{acmart}

\usepackage[T1]{fontenc}
\usepackage[utf8]{inputenc}
\usepackage{fontawesome5}

\usepackage{mathpartir}
\usepackage{graphicx}
\usepackage{hyperref}
\usepackage{multicol}
\usepackage{stmaryrd} %
\usepackage{xspace}
\usepackage{xcolor}
\usepackage{amsfonts}
\usepackage{adjustbox}
\usepackage{soul}
\usepackage{graphicx}

\usepackage{minted}
\newmintinline[ocamli]{ocaml}{}

\usepackage[nameinlink,noabbrev]{cleveref}
\crefname{section}{\S\!}{\S}
\Crefname{section}{Section}{Sections}
\crefname{figure}{Figure}{Figures}
\crefname{theorem}{Theorem}{Theorems}

\newtheorem{theorem}{Theorem}[section]

\usepackage{iris}
\newcommand{\captionlabel}[2]{%
  \vspace{-0.8em}%
  \caption{#1}%
  \Description{#1}%
  \label{#2}
  \vspace{-0.8em}%
}

\renewcommand{\TirNameStyle}[1]{\hypertarget{#1}{\textsc{#1}}}

\newcommand{\lang}{ParLang\xspace}
\newcommand{\logic}{Parcas\xspace}
\newcommand{\heaplang}{HeapLang\xspace}

\newcommand{\Values}{\mathcal{V}}
\newcommand{\vals}{\vec{\val}}
\newcommand{\vunit}{()}
\newcommand{\loc}{\ell}
\renewcommand{\Loc}{\mathcal{L}}
\newcommand{\vint}{i}
\newcommand{\vbool}{b}
\newcommand{\vtrue}{\mathsf{true}}
\newcommand{\vfalse}{\mathsf{false}}
\newcommand{\vfun}[3]{\hat\mu#1\,#2.\,#3}

\newcommand{\primitive}{\bowtie}
\newcommand{\symbeq}{\mathord{==}}
\newcommand{\symbor}{\lor}
\newcommand{\symband}{\land}
\newcommand{\sequence}{\mathord{;}}

\newcommand{\elet}[3]{\mathsf{let}\,#1\,=\,#2\,\mathsf{in}\,#3}
\newcommand{\efun}[3]{\mu#1\,#2.\,#3}
\newcommand{\efunnonrec}[2]{\lambda#1.\,#2}

\newcommand{\ecall}[2]{#1\,#2}
\newcommand{\epar}[2]{#1\,||\,#2}

\newcommand{\ealloc}[2]{\mathsf{alloc}\;#1\;#2}
\newcommand{\eload}[2]{#1[#2]}
\newcommand{\estore}[3]{#1[#2]\!\leftarrow\! #3 }
\newcommand{\elength}[1]{\mathsf{length}\,#1}
\newcommand{\eif}[3]{\mathsf{if}\,#1\,\mathsf{then}\,#2\,\mathsf{else}\,#3}
\newcommand{\ecas}[4]{\mathsf{CAS}\,#1\,#2\,#3\,#4}
\newcommand{\eeq}[2]{#1\,\symbeq\,#2}
\newcommand{\blockrepeat}[2]{#2^{#1}}
\newcommand{\eapp}[2]{#1\,#2}
\newcommand{\ecallprim}[2]{#1\primitive#2}

\newcommand{\etick}{\textsf{tick}}

\newcommand{\defineq}{\mathrel{::=}}

\newcommand{\store}{\sigma}
\renewcommand{\subst}[3]{[#2/#1]#3}
\renewcommand{\dom}[1]{\mathsf{dom}(#1)}
\newcommand{\singlemap}[2]{\{[#1:=#2]\}}
\newcommand{\blockupd}[3]{[#2:=#3]#1}
\newcommand{\wal}{w}
\newcommand{\wals}{\vec\wal}
\newcommand{\length}[1]{|#1|}

\newcommand{\ectx}{K}
\newcommand{\llangle}{{\langle\kern-0.25em\langle}}
\newcommand{\rrangle}{\rangle\kern-0.25em\rangle}
\newcommand{\efillctx}[2]{#1\llangle#2\rrangle}
\newcommand{\khole}{\square}

\newcommand{\minispace}{\kern1pt}
\newcommand{\sepconfig}{\minispace{}/\minispace{}}
\newcommand{\config}[2]{#1\sepconfig#2}

\newcommand{\petit}[1]{\text{\scriptsize\rm\sf#1}}
\newcommand{\steprightarrowrule}[1]
{\mathrel{\xrightarrow{\adjustbox{margin=0ex 0ex 0ex 1.2ex}{\raisebox{-0.2ex}[0ex][0ex]{#1}}}}}
\newcommand{\purestep}[2]{#1 \;\steprightarrowrule{\petit{pure}}\; #2}
\newcommand{\headstep}[4]{\config{#1}{#2} \,\steprightarrowrule{\petit{head}}\, \config{#3}{#4}}
\newcommand{\step}[2]{#1 \;\longrightarrow\; #2}
\newcommand{\steprtcname}{\longrightarrow^\star}
\newcommand{\steprtc}[2]{#1 \;\steprtcname\; #2}

\newcommand{\prompt}[2]{#1\;#2}
\newcommand{\wpg}[3]{\textsf{wp}\;\prompt{#1}{#2}\;\{#3\}}
\newcommand{\wpgex}[5]{\wpg{#1}{#2}{\lambda\,#3\,#4.\,#5}}
\newcommand{\wpglong}[3]{\textsf{wp}\;\prompt{#1}{#2}\;\left\{#3\right\}}
\newcommand{\wpgexlong}[5]{\wpglong{#1}{#2}{\lambda\,#3\,#4.\,#5}}
\newcommand{\wpgen}[3]{\textsf{wpg}\;\prompt{#1}{#2}\;\{#3\}}
\newcommand{\stateinterpname}{\textsf{interp}\xspace}
\newcommand{\stateinterp}[3]{\stateinterpname\,#1\,#2\,#3}

\newcommand{\mapm}{\textdom{Map}}

\renewcommand{\star}{\ast}
\newcommand{\wider}[1]{\,#1\,}
\newcommand{\morespacingaroundstar}{%
\let\oldstar\star
\renewcommand{\star}{\wider\oldstar}%
}
\newcommand{\morespacingaroundwedge}{%
\let\oldwedge\wedge
\renewcommand{\wedge}{\wider\oldwedge}%
}
\newcommand{\pure}[1]{\ulcorner #1 \urcorner}

\newcommand{\iTrue}{\top}
\newcommand{\pre}{\varphi}
\newcommand{\post}{\psi}

\newcommand{\qp}{q}
\newcommand{\ofs}{i}

\newcommand{\bigast}[2]{{\scaleobj{2}{\ast}}_{#1}\,#2}
\newcommand{\bigsepnbase}[4]{\bigast{#1[#2;#3)}{#4}}
\newcommand{\bigsepn}[4]{\bigsepnbase{#1 \in}{#2}{#3}{#4}}

\renewcommand{\state}{S}
\newcommand{\smallconfig}[3]{#1\sepconfig#2\sepconfig#3}
\newcommand{\phruple}[3]{#1\sepconfig#2\sepconfig#3}
\newcommand{\thruple}[3]{(#1,#2,#3)}

\newcommand{\vertex}{t}

\newcommand{\vertexs}{\mathcal{L}}
\newcommand{\tasktree}{T}
\newcommand{\tleaf}[1]{#1}
\newcommand{\tpar}[2]{#1\otimes#2}

\newcommand{\parastep}[2]{#1 \;\steprightarrowrule{\petit{sched}}\; #2}
\newcommand{\graph}{g}
\newcommand{\tickmap}{c}
\newcommand{\vertices}[1]{\textsf{vertices}(#1)}
\newcommand{\leaves}[1]{\textsf{leaves}(#1)}
\newcommand{\graphedge}[2]{(#1,#2)}

\newcommand{\littlenode}[2]{\tpar{\tleaf{#1}}{\tleaf{#2}}}

\newcommand{\complex}[1]{\mathcal{O}(#1)}
\newcommand{\workc}[1]{\mathcal{W}(#1)}
\newcommand{\spanc}[2]{\mathcal{S}(#1,#2)}
\newcommand{\scanwk}{\textsf{scan}^{eq}_{W}}
\newcommand{\scansp}{\textsf{scan}^{eq}_{S}}

\newcommand{\funcname}{f}
\newcommand{\body}{\expr}
\newcommand{\argname}{x}
\newcommand{\allocsize}{n}

\newcommand{\kw}[1]{\mathsf{#1}}
\newcommand{\roundup}[1]{\lceil #1 \rceil}
\newcommand{\logtwoup}[1]{\roundup{\log_2 #1}}
\newcommand{\parforname}{\textsf{parfor}\xspace}
\newcommand{\parfor}[3]{\parforname\,#1\, #2\,#3}
\newcommand{\selfname}{f}
\newcommand{\lowbound}{a}
\newcommand{\highbound}{b}
\newcommand{\diffname}{(\highbound - \lowbound)}
\newcommand{\midname}{mid}
\newcommand{\parforarg}{h}
\newcommand{\zero}{0}
\newcommand{\one}{1}
\newcommand{\mpost}{Q}

\newcommand{\tabulatename}{\textsf{tabulate}\xspace}
\newcommand{\tabulate}[2]{\tabulatename\,#1\,#2}

\newcommand{\workboundname}{\textsf{WorkBound}\xspace}
\newcommand{\workbound}[2]{\workboundname\,#1\,#2}
\newcommand{\spanboundname}{\textsf{SpanBound}\xspace}
\newcommand{\spanbound}[3]{\spanboundname\,#1\,#2\,#3}

\newcommand{\scanname}{\textsf{scan}\xspace}
\newcommand{\scanned}[2]{\textsf{scanned}\,#1\,#2}
\newcommand{\scanarg}{s}

\newcommand{\slengthname}{\textsf{s\_length}\xspace}
\newcommand{\slength}[1]{\slengthname\,#1}
\newcommand{\sloadname}{\textsf{s\_load}\xspace}
\newcommand{\sload}[2]{\sloadname\,#1\,#2}
\newcommand{\sstorename}{\textsf{s\_store}\xspace}
\newcommand{\sstore}[3]{\sstorename\,#1\,#2\,#3}
\newcommand{\ssplitname}{\textsf{s\_split}\xspace}
\newcommand{\ssplit}[3]{\ssplitname\,#1\,#2\,#3}

\newcommand{\sortname}{\textsf{sort}\xspace}
\newcommand{\sortwk}{\textsf{sort}_W}
\newcommand{\sortsp}{\textsf{sort}_S}

\newcommand{\mergename}{\textsf{merge}\xspace}
\newcommand{\mergeop}[2]{\textsf{puremerge}(#1, #2)}
\newcommand{\scopyseqname}{\textsf{copy\_seq}\xspace}
\newcommand{\scopyseq}[2]{\scopyseqname\,#1\,#2}
\newcommand{\smergeseqname}{\textsf{merge\_seq}\xspace}
\newcommand{\smergeseq}[3]{\smergeseqname\,#1\,#2\,#3}
\newcommand{\binsearchname}{\textsf{binsearch}\xspace}
\newcommand{\binsearch}[2]{\binsearchname\,#1\,#2}
\newcommand{\mergewk}{\textsf{merge}_W}
\newcommand{\mergesp}{\textsf{merge}_S}
\newcommand{\mergewkeq}{\textsf{merge}^{eq}_W}
\newcommand{\mergespeq}{\textsf{merge}^{eq}_S}
\newcommand{\sliceown}[2]{\textsf{slice}\,#1\,#2}
\newcommand{\sorted}[1]{\textsf{sorted}\,#1}

\newcommand{\stackcreatename}{\textsf{create}\xspace}
\newcommand{\stackpushname}{\textsf{push}\xspace}
\newcommand{\stackpush}[2]{\stackpushname\,#1\,#2}
\newcommand{\stackpopname}{\textsf{pop}\xspace}
\newcommand{\stackpop}[1]{\stackpopname\,#1}
\newcommand{\stackownname}{\textsf{stack}}
\newcommand{\stackown}[4]{\stackownname\,#1\,#2\,#3\,#4}
\newcommand{\playername}{\textsf{participant}\xspace}
\newcommand{\player}[2]{\playername\,#1\,#2}

\newcommand{\sumall}[1]{\textsf{sumall}(#1)}
\newcommand{\ispath}[5]{#1,#2\vdash #3 \rightsquigarrow_{#4} #5}
\newcommand{\reducible}[3]{\textsf{AllRed}\,#1\,#2\,#3}
\newcommand{\safe}[3]{\textsf{Safe}\,#1\,#2\,#3}

\newcommand{\comptreename}{\textsf{CompTree}}
\newcommand{\comptree}[2]{\comptreename\,#1\,#2}
\newcommand{\hasnoloop}[1]{\textsf{has\_no\_loop}\,#1}
\newcommand{\pureinvname}{\textsf{pureinv}}
\newcommand{\pureinv}[4]{\pureinvname\,#1\,#2\,#3\,#4}
\newcommand{\interpresname}{\textsf{interp\_res}\xspace}
\newcommand{\interpres}[3]{\interpresname\,#1\,#2\,#3}
\newcommand{\heapinterp}[1]{\textsf{heap}\,#1}
\newcommand{\initspan}[2]{\textsf{init\_span}\,#1\,#2}
\newcommand{\boundedsrcname}{\textsf{cascade}\xspace}
\newcommand{\boundedsrc}[2]{\boundedsrcname\,#1\,#2}
\newcommand{\tracktransfer}[1]{\textsf{forks\_and\_joins}\,#1}
\newcommand{\interpwork}[1]{\textsf{interp\_work}\,#1}
\newcommand{\sources}[1]{\textsf{sources}(#1)}

\newcommand{\cantransfername}{\textsf{transferable}\xspace}
\newcommand{\cantransfer}[2]{\cantransfername\,#1\,#2}
\newcommand{\isforkname}{\textsf{is\_fork}}
\newcommand{\isfork}[3]{\isforkname\,#1\,#2\,#3}
\newcommand{\isjoinname}{\textsf{is\_join}}
\newcommand{\isjoin}[3]{\isjoinname\,#1\,#2\,#3}
\newcommand{\spaninitial}{\tickmap_0}

\newcommand{\interpspan}[2]{\textsf{interp\_span}\,#1\,#2}
\newcommand{\initwork}[1]{\textsf{init\_work}\,#1}
\newcommand{\authwork}[1]{\textsf{auth\_work}\,#1}
\newcommand{\authspan}[1]{\textsf{auth\_span}\,#1}
\newcommand{\updatecell}[3]{\textsf{incr}\,#1\,#2\,#3}
\newcommand{\taskids}{\textsf{TaskIds}}

\newcommand{\disjoint}[2]{#1\,\cap\,#2 = \emptyset}

\newcounter{remark}[section]

\usepackage{xspace}
\usepackage{expl3}
\ExplSyntaxOn
\newcommand\latinabbrev[1]{
  \peek_meaning:NTF . {%
    \emph{#1}\@}%
  { \peek_catcode:NTF a {%
      \emph{#1}.\@\xspace}%
    {\emph{#1}.\@\xspace}}}
\ExplSyntaxOff

\def\ie{\latinabbrev{i.e}}

\setcopyright{cc}
\setcctype{by}
\acmDOI{10.1145/3828679}
\acmYear{2026}
\acmJournal{PACMPL}
\acmVolume{10}
\acmNumber{ICFP}
\acmArticle{281}
\acmMonth{8}
\acmSubmissionID{icfp26main-p30-p}
\received{2026-03-01}
\received[accepted]{2026-05-13}

\renewcommand{\TirNameStyle}[1]{\hypertarget{#1}{\textsc{#1}}}
\newcommand{\RULE}[1]{\hyperlink{#1}{\textsc{#1}}\xspace}

\makeatletter
\def\arcr{\@arraycr}
\makeatother

\setcitestyle{nosort}

\begin{document}

\title{A Separation Logic for Parallel Time Complexity with Work and Span Credits}

\author{Alexandre Moine}
\orcid{0000-0002-2169-1977}
\affiliation{%
  \institution{New York University}
  \city{New York}
  \country{USA}
}
\email{alexandre.moine@nyu.edu}

\author{Sam Westrick}
\orcid{0000-0003-2848-9808}
\affiliation{%
  \institution{New York University}
  \city{New York}
  \country{USA}
}
\email{shw8119@nyu.edu}

\author{Joseph Tassarotti}
\orcid{0000-0001-5692-3347}
\affiliation{%
  \institution{New York University}
  \city{New York}
  \country{USA}
}
\email{jt4767@nyu.edu}
\authornote{Also affiliated with Amazon Web Services. This paper does not reflect the views of Amazon Web Services.}

\begin{abstract}
We present Parcas, a concurrent separation logic for verifying the
parallel time complexity of fork-join programs.
In order to abstract from the specifics of the machine,
time complexity for parallel programs is given
in terms of two metrics: the work, measuring
the total number of operations, and the span, measuring
the longest chain of sequential dependencies.
Together, these two metrics determine the running time
on any number of processors.
For proving bounds on the work and span,
Parcas is equipped with work credits and span credits,
logical devices that represent permissions to incur costs.

Work credits are a straightforward adaptation of time credits,
a standard tool for bounding time complexity of sequential programs,
and can be split additively between parallel tasks.
Span credits, however, require a fundamentally different treatment.
Indeed, the span of the parallel composition of two tasks
is the maximum of the span of the two tasks.
To account for this,
we propose a rule for duplicating span credits
at fork points, with each copy tagged by a logical task identifier
that restricts which task may spend them.
A transfer rule allows unused span credits to be
forwarded across sequential compositions to subsequent tasks.
The logic is expressive enough to give modular, higher-order
specifications for common parallel primitives such
as a parallel for loop and a tabulate function.
We demonstrate Parcas on several case studies, including
parallel prefix sums, parallel merge sort, and
a variant of Treiber's lock-free stack that mixes concurrency
with parallelism.
All the presented results are mechanized in the Rocq prover
using the Iris separation logic framework.

\end{abstract}



\maketitle

\section{Introduction}
\label{sec:intro}
A variety of techniques exist for formally establishing bounds on the resource consumption of programs.
One approach is to introduce substructural \emph{credits} or \emph{potentials} into a type system~\citep{hofmann-jost-03} or program logic~\citep{atkey-11}, which must be spent at each point in which a program consumes a corresponding resource.
The credits thereby represent a ``budget'' that upper bounds the resource consumption of a program.
This approach has been used to reason about a range of different kinds of resources, including running time~\citep{hoffmann-aehlig-hofmann-raml-12, chargueraud-pottier-uf-sltc-19, mevel-jourdan-pottier-19}, heap memory under both manual management~\citep{hofmann-jost-03} and garbage collection~\citep{niu-hoffmann-18, madiot-pottier-22, moine-chargueraud-pottier-23}, and stack space~\citep{campbell-09, carbonneaux-14}.

This paper introduces \logic{}, a concurrent separation logic with credits for bounding running time cost in fork-join parallel programs.
We consider programs written in a parallel functional language, similar in style to
NESL~\citep{blelloch-nesl-93} and parallel ML-family languages (such as
Manticore~\cite{DBLP:conf/icfp/FluetRRS08},
MaPLe~\cite{mpl-2020},
multicore OCaml~\cite{DBLP:journals/pacmpl/Sivaramakrishnan20},
and
Futhark~\cite{Henriksen:2017:FPF:3062341.3062354}).
For such programs, running time cost can be accounted for in terms of two metrics: the \emph{work}, which measures the total number of operations that must be performed, and the \emph{span} (or depth), which measures the longest chain of dependent operations.
This latter metric is important because it limits how much speed-up can be obtained from increasing the number of parallel processors used to execute the program.
More precisely, when a parallel program with work $w$ and span $s$ is executed on $p$ processors using an appropriate scheduling algorithm, the running time will be $\complex{\max(w/p, s)}$ by Brent's theorem~\citep{Brent74, blumofe-leiserson-99}, and this bound is optimal up to constant factors.
Thus, analyzing both work and span is important to get a complete picture of a parallel program's execution costs.

For reasoning about work in parallel programs, \logic{} uses a relatively straightforward adaptation of prior logics' techniques for representing and tracking credits.
Since the total work is the sum of the work incurred by each parallel thread, one can additively split credits for work between threads, which is easily represented in separation logic.
And, if a thread consumes fewer work credits than its budget allows, we can combine back together the excess credits at the join point.

However, reasoning about span is fundamentally different, because the total span of a parent thread in the fork-join model is the \emph{maximum} of the span of its children.
Thus, additively splitting credits across children is not sufficient, and it would be unsound to naively combine together span credits at the join point.
While there has been prior work developing a substructural type system using potentials to bound work and span in a language similar to the one we consider~\citep{hoffmann-shao-parallel}, that work restricts to first-order pure programs. Moreover, the approach used in that work exploits certain constraints of the type system, which are challenging to translate to the setting of a program logic without imposing severe restrictions on the expressivity of the logic.
Other works have developed type systems with sized types for analyzing a form of parallel complexity in the Pi-calculus~\citep{baillot-ghyselen-24}, session typed languages~\citep{das-parallel-session}, or interaction nets~\citep{gimenez-moser-16}, but these execution models are quite different from the data-parallel functional language with shared mutable state that we consider here.

\logic{} addresses the challenge of reasoning about span by tagging span credits with a logical \emph{task identifier}, which restricts which threads may spend the credit.
When a parent thread forks, it creates duplicate copies of its span credits, one set for each child, with updated identifiers that restrict each child's set of credits to be spent only by that child.
Additionally, the logic provides a proof rule to promote the identifier associated with a credit, allowing unused credits to be transferred and then spent by other threads, so long as there is an appropriate ordering between them
in the computation graph.
Apart from credits, \logic{} is a standard separation logic.

We show that \logic{} is sound: if a program is verified with
$w$ initial work credits and $s$ initial span credits, then
it is always safe (no execution ever gets stuck), and the work
and span of every execution are bounded by $w$ and $s$, respectively.
Here, the work and span of individual executions are given in terms of
a graph structure called a \emph{computation graph} (also known as a
\emph{cost graph}~\cite{DBLP:books/cu/Ha2016}), and \logic{} provides worst-case bounds on
the work and span of such graphs.
We note that this approach is sufficient for the applicability of
scheduling theorems (such as Brent's theorem discussed above), because
every execution under a particular scheduler yields a computation graph within
our model, and \logic{} gives bounds over all such graphs.

By combining traditional work credits with these new span credits, \logic{} is expressive enough to establish asymptotically tight bounds for a number of challenging parallel algorithms.
Moreover, because the logic is higher-order, we are able to give modular, parametric specifications for higher-order parallel operations such as a parallel for loop and a tabulate function.
We demonstrate \logic{} by verifying cost bounds for a number of examples, including parallel prefix sum, merge sort, and a variant of Treiber's stack, which mixes concurrency with parallelism.

All of the results in this paper have been mechanized in the Rocq proof assistant on top of the Iris separation logic framework~\citep{mechanization}.

\section{Key Ideas}
\label{sec:key_ideas}
In this section,
we present the key ideas of our approach.
First, we describe the parallel execution model we consider~(\cref{sec:key:dag}).
Then, we present a small example illustrating
the notions of work and span~(\cref{sec:key:example}).
Next, we present work and span credits as well as the weakest precondition of \logic~(\cref{sec:key:weakestpre}).
We show the key reasoning rules of work and span credits~(\cref{sec:key:credits})
and conclude with a proof sketch of the verification of our small example~(\cref{sec:key:example_verif}).

\subsection{Computation Graphs, Work, and Span}
\label{sec:key:dag}

We consider programs in an ML-like language
with a parallel primitive~$\epar{\expr_1}{\expr_2}$.
The full formal semantics of the language is given later in \Cref{sec:syntax_and_semantics},
but for now, the high-level idea is that the expression~$\epar{\expr_1}{\expr_2}$
evaluates
the two expressions $\expr_1$ and $\expr_2$ in parallel,
waits for them to terminate their executions
on values $\val_1$ and $\val_2$, respectively, and
returns an array of size two storing $\val_1$ and $\val_2$.
Parallel primitives can be arbitrarily nested.

In order to keep track of the parallel structure
of the program, the semantics uses a \emph{computation graph}~\citep{nesl-proof}.
A vertex of the computation graph
is called a \emph{task}
and represents a sequential unit of computation.
Every task is annotated with a unique identifier $\vertex \in \taskids$.
Edges between tasks represent a dependency.
Two operations add edges to the computation graph.
First, when a task~$\vertex$ starts executing a parallel
expression $\epar{\expr_1}{\expr_2}$,
it \emph{forks}.
To reflect this operation,
we generate two fresh tasks $\vertex_1$ and $\vertex_2$,
and add edges $(\vertex,\vertex_1)$ and $(\vertex,\vertex_2)$
to the computation graph.
Second, when $\vertex_1$ and $\vertex_2$ finish their execution,
they \emph{join}.
To reflect this operation,
we generate a fresh task $\vertex'$ for the continuation,
and add edges $(\vertex_1,\vertex')$ and $(\vertex_2,\vertex')$
to the computation graph.
The computation graph is always acyclic.

To represent operations that have some cost to execute, our programming language
is equipped with a ghost operation ``$\etick$''
that incurs a cost of $1$ for the task executing it.
We reflect this cost by adding a \emph{weight} to each vertex in the
computation graph, reflecting the number of tick operations the task executed.
We can then formally define
the \emph{work} as the total weight, that is,
the sum of the weight of every vertex.
The \emph{span} is defined as
the heaviest path (or \emph{critical path}),
where the weight of a path is the sum of
the weights of its vertices.
By instrumenting a program appropriately with $\etick$, we can track
different notions of cost depending on the application under consideration.
For example, for a sorting algorithm, we might add a tick only after each comparison,
so that the cost incurred by the program corresponds to the number of comparisons.

Equipped with these weights, our
computation graphs form a cost model (the \emph{binary fork-join model}),
commonly used for analyzing parallel programs.
Because our language has mutable state that can be shared by threads,
different interleavings of thread steps can give rise to different
computation graphs. When analyzing the cost of a program,
we will upper bound the worst-case cost across all of these possible graphs.
A worst-case analysis on the work and span
allows us to give an upper bound on the asymptotic running
time on other parallel cost models, such as PRAM, when the program is executed using an appropriate scheduler~\citep{blelloch-et-al-20}.
It is important to consider the worst case across all computation graphs for a program, because some graphs may have
an unrealistically small cost due to a particular ordering of load and store operations.

\subsection{A Small Example}
\label{sec:key:example}
\begin{figure*}
\begin{minipage}{.5\textwidth}
\small
\begin{minted}[escapeinside=!!]{ocaml}
let f (x: int) : int =
  (if x <= 3
   then (!$\underbrace{\texttt{\colorbox{lightgray}{tick}{\color{gray}; ...}}}_{2k\ \text{ticks total}}$!) else (!$\underbrace{\texttt{\colorbox{lightgray}{tick}{\color{gray}; ...}}}_{k\ \text{ticks total}}$!)); x

let example =
  let a = f 1
  let left() = f 2
  let right() = (!{\colorbox{lightgray}{tick}}!; (f 3 || f 4))
  let (b, (c, d)) = (left() || right())
  let e = f 5
  in a + b + c + d + e
\end{minted}
\end{minipage}
\hfill
\begin{minipage}{.48\textwidth}
\includegraphics[width=2.5in]{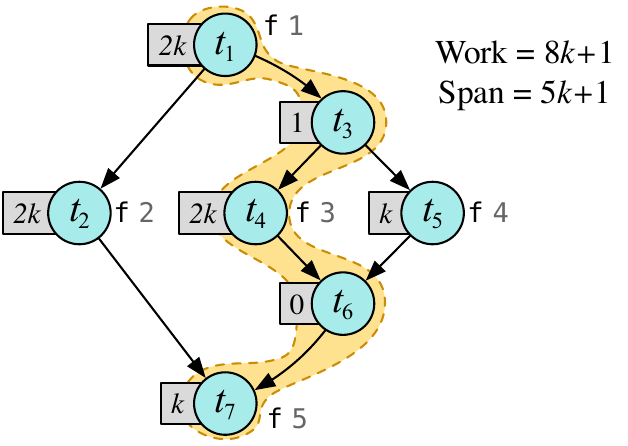}
\end{minipage}
\captionlabel{Small example program and its computation graph. The function
\ocamli{f} performs either $k$ or $2k$ sequential ticks, depending
on its argument; these ticks occur sequentially (no use of \ocamli{||} inside \ocamli{f}).
Vertices in the computation graph ($t_1, t_2, \ldots$) are annotated with
their weights (number of ticks executed). The work
(total number of ticks executed) is $8k+1$. The span (weight of the
critical path, highlighted in the graph) is $5k+1$.}{fig:example}
\end{figure*}

In order to illustrate the notions
of computation graph, work, and span,
\Cref{fig:example} presents a small example.
The example program
makes use of a function
\ocamli{f} from \ocamli{int} to \ocamli{int},
which incurs sequential cost.
Executing \ocamli{f} performs either $2k$ ticks
if its argument is less than or equal to $3$,
or $k$ ticks otherwise, with $k$ being an arbitrary constant.
The function \ocamli{f}
is purely sequential and illustrates the fact that
the runtime cost of a function can depend on the
value of its arguments.

The example program first calls \ocamli{f 1}.
It then defines two functions: the \ocamli{left} function, which calls \ocamli{f 2},
and the \ocamli{right} function, which
itself executes a tick, and
calls in parallel \ocamli{f 3} and \ocamli{f 4}.
The program executes in parallel \ocamli{left()} and \ocamli{right()},
and finally calls \ocamli{f 5} before returning the sum of all results.
The computation graph, drawn on the right of \Cref{fig:example},
illustrates the parallel structure of the program.
Vertices are annotated with their identifiers,
and on the right with their weights (number of ticks executed).
By counting the calls to \ocamli{f}
and the ticks they produced, we can see that
the total work is $8k+1$.
For the span, we need to identify the heaviest path.
The heaviest path of this program is the one highlighted in the figure,
which has a weight of $5k+1$.

\subsection{Work and Span Credits and the Weakest Precondition}
\label{sec:key:weakestpre}

\paragraph{Credits}
Our logic employs two kinds of credits:
\emph{work credits} to account for work
and \emph{span credits} to account for span.
First, $n$ \emph{work credits}, with $n \in \mathbb{N}$ and written $\workc{n}$,
allow for executing a total of $n$ $\etick$ operations.
Work credits can be distributed freely between tasks.
Second, $n$ \emph{span credits}, with $n \in \mathbb{N}$ and written $\spanc{\vertex}{n}$ for some task $\vertex$,
allow for executing multiple $\etick$ operations, perhaps in parallel,
which in all contribute no more than weight $n$ to the heaviest path from $\vertex$.
Both work credits and span credits can be split and combined, as shown by the
following rules.
\begin{mathpar}\morespacingaroundstar
\inferrule[WorkSplit]{}
{\workc{n+m} \equiv \workc{n} \star \workc{m}}

\inferrule[SpanSplit]{}
{\spanc{\vertex}{n+m} \equiv \spanc{\vertex}{n} \star \spanc{\vertex}{m}}
\end{mathpar}

Work credits can always be freely split and combined and are not
associated with any particular task.
This aligns with the intuition that work credits only serve to count the
total number of tick operations, regardless of which task executes them.
For span credits, $\spanc{\vertex}{n}$ allows only the task~$\vertex$ to
execute $n$ $\etick$ operations.
This restriction is necessary to ensure that span credits can be associated
with paths in the computation.
As we will see, span credits can be soundly transferred across subexpression
evaluation, which in graph terms corresponds to transferring credits
from a task~$t$ to some other task $t'$ that is a member of
all paths originating from $t$.
For example, span credits can be transferred
from a parent task about to fork two children to the task joining them
(recall that a fork operation creates two new tasks, one for each child,
and a join operation creates a new task for the continuation).
This notion of ``transfer'' is
encoded within the logic as a key reasoning rule,
which we discuss in more detail in \Cref{sec:key:credits}.

\paragraph{The weakest precondition}
In order to relate work and span credits to the execution of a program,
we introduce \logic.
\logic is built on Iris~\citep{iris}, and we follow its syntax.
We denote an Iris assertion by $\pre$,
a separating conjunction by $\pre \star \pre'$,
and a separating implication (magic wand) by $\pre \wand \pre'$.
A meta-logical proposition~$U$ is called \emph{pure} and written $\pure{U}$.
Assertion equivalence is denoted~$\pre \equiv \pre'$.
Central to our logic is the weakest precondition~(WP) modality:
\[\wpgex\vertex\expr{\vertex'}\val{\pre}\]
Compared to the standard Iris WP,
this WP accounts for task identifiers,
in a similar manner as \citet{dislog}.
Here,~$\vertex$ represents the identifier of the task executing~$\expr$,
which we refer to as the~\emph{current task}.
The postcondition has the form
$\lambda\,\vertex'\,\val.\; \pre$, binding two variables in $\pre$:
the \emph{end task}~$\vertex'$ and the result value~$\val$.
The end task identifier characterizes the task when the computation is completed.
The current and end task identifiers need not coincide:
in the \ocamli{example} program from \Cref{sec:key:example},
at the time of executing the program,
the current identifier is $\vertex_1$ and the end identifier is $\vertex_7$.
When the details of the bound variables in the postcondition can be omitted, we abbreviate it using the metavariable $\post$.

The formal statement of the soundness theorem of \logic will be presented later in \Cref{sec:soundness_thm},
but intuitively, it says that if
the user verified a program $\expr$ with $w$ initial work credits
and $s$ initial span credits,
that is, if $\workc{w} \star \spanc{\vertex_0}{s} \vdash \wpgex{\vertex_0}{\expr}{\_}{\_}{\iTrue}$ holds for some initial task $\vertex_0$,
then
(1) the work of $\expr$ is bounded by $w$,
(2) the span of $\expr$ is bounded by $s$, and
(3) $\expr$ is safe (that is, never gets stuck, a standard property guaranteed by separation logic).
Note that when applying the soundness theorem, we fix the initial work and span credits that will be available to verify $\expr$ up front---
there is no way to create additional credits during the proof.

\subsection{Key Reasoning Rules for Credits}
\label{sec:key:credits}
\begin{figure}
\centering\small\morespacingaroundstar
\begin{mathpar}
\inferrule[Tick]
{\workc{1} \\ \spanc{\vertex}{1}}
{\wpgex{\vertex}{\etick}{\_}{\val}{\pure{\val = \vunit}}}

\inferrule[Transfer]
{\spanc{\vertex}{n} \\
  \wpgex{\vertex}{\expr}{\vertex'}{\val}{\spanc{\vertex'}{n} \wand \post\,\vertex'\,\val}}
{\wpg{\vertex}{\expr}{\post}}

\inferrule[Par]
{\spanc{\vertex}{n} \\
  \forall \vertex_1\,\vertex_2.\;
  \spanc{\vertex_1}{n} \star \spanc{\vertex_2}{n} \wand
  \wpg{\vertex_1}{\expr_1}{\post_1} \star \wpg{\vertex_2}{\expr_2}{\post_2}}
{\wpgex{\vertex}{(\epar{\expr_1}{\expr_2})}{\_}{\val}
{\exists \loc\,\vertex_1\,\vertex_2\,\val_1\,\val_2.\;
\pure{\val = \loc} \star \loc \mapsto [\val_1;\val_2] \star \post_1\,\vertex_1\,\val_1 \star \post_2\,\vertex_2\,\val_2}}
\end{mathpar}
\captionlabel{Reasoning rules for credits, and for the tick and the parallel primitives}{fig:credits}
\end{figure}

\Cref{fig:credits}
presents reasoning rules associated with credits---%
the other reasoning rules of \logic are mostly standard and presented in \Cref{sec:rules}.
We present reasoning rules as inference rules where premises
are implicitly joined by~$\star$, and the horizontal bar stands for entailment.

\paragraph{Tick and Par}

\RULE{Tick} shows that a tick operation consumes both a work credit
and a span credit, with a matching identifier.
This follows from the fact that the $\etick$ operation incurs a cost of $1$ for the task executing it.

\RULE{Par} constitutes the crux of our approach.
At first sight, \RULE{Par} has the ingredients
of the standard separation logic rule for a parallel composition:
to verify that $\epar{\expr_1}{\expr_2}$ is correct,
the user has to verify~$\expr_1$ and $\expr_2$ separately.
\RULE{Par} differs from the standard rule in two ways.
First, the rule is indexed by the current task $\vertex$,
and universally quantifies over the identifiers $\vertex_1$ and $\vertex_2$
of the sub-tasks.
Second, the precondition of \RULE{Par}
consumes span credits $\spanc{\vertex}{n}$
and ``duplicates'' them for the two generated tasks $\vertex_1$ and $\vertex_2$.
Indeed, the precondition
requires the user to verify the WP of~$\expr_1$ at some identifier~$\vertex_1$
and of $\expr_2$ at some identifier $\vertex_2$,
while having access to span credits
$\spanc{\vertex_1}{n}$ and $\spanc{\vertex_2}{n}$.
The postcondition of \RULE{Par} reflects that
the parallel primitive returns an array containing the result of
the two evaluated expressions, and that their respective postconditions hold.

Why is this ``duplication'' of span credits sound?
The key is that span credits cannot be transferred between unrelated tasks.
When the parent task $\vertex$ (with a budget $\spanc{\vertex}{n}$) forks into
two child tasks~$\vertex_1$ and $\vertex_2$,
the two children are given two separate budgets $\spanc{\vertex_1}{n}$ and
$\spanc{\vertex_2}{n}$, and these two budgets cannot intermix.
\RULE{Par} therefore is able to independently bound the span of both $\expr_1$
and $\expr_2$ by $n$, which in turn bounds the span of
$(\epar{\expr_1}{\expr_2})$ by $n$ as well.

The reader might wonder why
\RULE{Par} does not mention the \emph{work} credits needed
for the two expressions $\expr_1$ and $\expr_2$.
The reason is that,
because we are working in a weakest
precondition calculus, any work credits needed for executing the two expressions
can be included as part of the proof of the second premise.
For example, suppose that we have previously proved a specification
for $\expr_i$ (for $i \in \{1,2\}$) of the
form $\forall \vertex_i.\;\workc{K} \wand \spanc{\vertex_i}{C} \wand \wpg{\vertex_i}{\expr_i}{\post_i}$,
which shows that $\expr_i$ requires $K$ work credits.
Then by combining this specification with $\workc{2 \times K}$,
and by using \RULE{WorkSplit},
we can derive $\forall \vertex_1\,\vertex_2.\;
\spanc{\vertex_1}{n} \star \spanc{\vertex_2}{n} \wand
\wpg{\vertex_1}{\expr_1}{\post_1} \star \wpg{\vertex_2}{\expr_2}{\post_2}$,
matching the second premise of \RULE{Par}.
In contrast, the span credits
need to appear explicitly in \RULE{Par}
because they must be transferred to the child threads.

\paragraph{The need for a transfer rule}

While \RULE{Par} allows a parent task to give span credits
to its children, there is no way for a child task
to give span credits back to its parent in a form they can use.
In particular, although~$\vertex_1$ and~$\vertex_2$ can put unused credits $\spanc{\vertex_1}{n_1}$ and $\spanc{\vertex_2}{n_2}$ into their postconditions $\post_1$ and $\post_2$,
which are passed back to the parent at the join point, the parent has no way to spend these credits or any leftover credits $\spanc{\vertex}{n'}$ it may have had from before the fork.
This is because the continuation of the parent task, after the join operation,
gets a fresh identifier $\vertex'$, so these credits with different identifiers cannot be used.
A restriction on only spending span credits with the appropriate tag is necessary to ensure soundness of the logic---since credits were duplicated as part of \RULE{Par}, we must be careful about which task can use them---but without some other mechanism, this restriction is too severe.
To see why, consider the following \RULE{Sequence} rule for a sequential composition, which is the same as the standard rule found in similar program logics, except for the addition of the task identifiers.\footnote{We derive \RULE{Sequence} from \RULE{Bind} and \RULE{LetVal} presented in \Cref{sec:rules}.}
\begin{mathpar}
\inferrule*[Left=Sequence]
{\wpgex{\vertex}{\expr_1}{\vertex'}{\val}{\wpg{\vertex'}{(\subst{x}{\val}{\expr_2})}{\post}}}
{\wpg{\vertex}{(\elet{x}{\expr_1}{\expr_2})}{\post}}
\end{mathpar}
This rule allows for verifying a sequential composition $\elet{x}{\expr_1}{\expr_2}$
by first verifying $\expr_1$ and then~$\expr_2$, in which we substitute $x$ with $\val$, the result of $\expr_1$.
However, notice the identifiers: $\expr_1$ is being verified at current task $\vertex$,
while $\expr_2$ is being verified at $\vertex'$, which is the end task of $\expr_1$.
These identifiers might potentially be different if $e_1$ contained a parallel composition.
As a result, with the rules we have discussed so far, if there are some leftover span credits $\spanc{\vertex}{n}$ after verifying $e_1$,
they cannot be used to verify $\expr_2$.

To solve this problem, we provide \RULE{Transfer}, which
allows for transferring span credits
from the current task to the end task.
Reading from bottom to top, the \RULE{Transfer} rule says:
if the current task is $\vertex$ executing $\expr$ with postcondition $\post$,
and the user has
$\spanc{\vertex}{n}$ span credits,
then it suffices to verify~$\expr$ but with a
postcondition enriched with $\spanc{\vertex'}{n}$ span credits,
where $\vertex'$ is the end task.

For some intuition as to why this rule is sound,
note that in the computation graph
(1) the task $\vertex$ \emph{dominates} the task $\vertex'$,
meaning that every path that passes through $\vertex'$ must pass through~$\vertex$, and
(2) the task $\vertex'$ is an active task, meaning that at this point in the proof,
it has no successor.
The first property ensures that
the heaviest path starting from~$\vertex'$ is a suffix of the heaviest
path starting from~$\vertex$,
and hence that the weight of the latter bounds the weight of the former.
The second property ensures that we don't have to
worry about possible paths starting from $\vertex'$ yet, there are none.
\Cref{fig:transfer-illustration} illustrates how the respective credits in \RULE{Transfer} cover different portions of the computation graph.
Combining \RULE{Sequence} with \RULE{Transfer}, we can easily prove the following
rule that exactly solves the problem with the initial version of \RULE{Sequence}.
\begin{mathpar}
\inferrule
{\spanc{\vertex}{n} \\ \wpgex{\vertex}{\expr_1}{\vertex'}{\val}{\spanc{\vertex'}{n} \wand \wpg{\vertex'}{(\subst{x}{\val}{\expr_2})}{\post}}}
{\wpg{\vertex}{(\elet{x}{\expr_1}{\expr_2})}{\post}}
\end{mathpar}
Here, the $\spanc{\vertex}{n}$ credits initially held by $\vertex$ are not needed for the execution of $\expr_1$, so they are transferred to use in verifying $\vertex'$. (The $\wand$ in the postcondition of the premise means that when proving the weakest precondition about $\subst{x}{\val}{\expr_2}$, we may assume access to $\spanc{\vertex'}{n}$ credits.)

The usage pattern for organizing span credits before executing a
parallel primitive
(or any library function that may call the parallel primitive)
is to split the available span credits in two assertions:
one will be given to \RULE{Par} so that children have span credits to spend,
and the other will be transferred to the continuation with \RULE{Transfer}.

\begin{figure*}
\begin{minipage}[b]{0.43\textwidth}
\includegraphics[width=\textwidth]{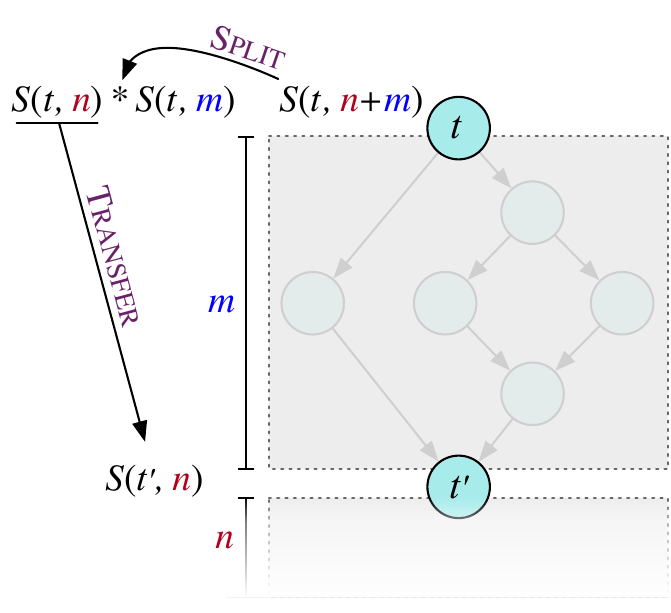}
\captionlabel{Illustration of using the \textsc{Transfer} rule to bound the
span of a region of the computation graph. All paths from $\vertex$ must pass
through $\vertex'$, so the weight of the heaviest path from $\vertex$ is the sum of the heaviest path in the grey region and the heaviest path from $t'$.}{fig:transfer-illustration}
\end{minipage}
\hfill
\begin{minipage}[b]{0.5\textwidth}
\includegraphics[width=\textwidth,trim=0 0.25in 0 0]{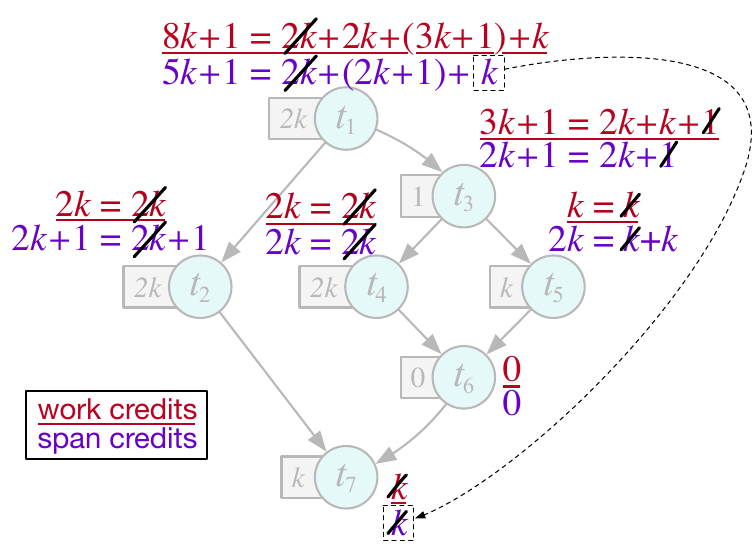}
\captionlabel{Accounting for all credits in the verification of
Figure~\ref{fig:example}.
The figure can be read from top to bottom: we start with $8k+1$ work
credits and $5k+1$ span credits at $\vertex_1$.
Work credits are shown in red (underlined), and span credits in
purple.
Crossed-out credits are consumed at the associated task. The
dashed arrow indicates a \textsc{Transfer}.}{fig:intro-example-credits}
\end{minipage}
\end{figure*}

\subsection{Verification of our Small Example}
\label{sec:key:example_verif}

With the rules presented in \Cref{sec:key:credits}
(as well as the other rules of \logic{} presented in \Cref{sec:rules}),
we can verify the work and span bounds of the program from \Cref{sec:key:example}.
First, we specify the auxiliary sequential function \ocamli{f} as follows:
\begin{mathpar}\morespacingaroundstar
\inferrule
{(\pure{\texttt{n} \leq 3} \star \workc{2k} \star \spanc{\vertex}{2k})
\;\;\lor\;\;
(\pure{\texttt{n} > 3} \star \workc{k} \star \spanc{\vertex}{k})}
{\wpgex\vertex{(\texttt{f}\,\texttt{n})}{\vertex'}{\val}{\pure{\vertex'=\vertex \wedge \val=\texttt{n}}}}
\end{mathpar}
In the above specification, note that the task $\vertex$
is (implicitly) universally quantified, meaning that this specification
can be used by any task.
Moreover, this specification enforces that \ocamli{f} is sequential:
the end task identifier $t'$ must be equal to the current task identifier $t$.
Second, we verify the main program as follows:
\begin{mathpar}\morespacingaroundstar
\inferrule
{\workc{8\times k+1} \star \spanc{\vertex_1}{5\times k+1}}
{\wpgex{\vertex_1}{(\texttt{example}\,\vunit)}{\_}{\val}{\pure{\val=15}}}
\end{mathpar}
The proof goes as follows, with \Cref{fig:intro-example-credits} illustrating the flow of credits throughout the argument.
First, we split work and span credits
and consume $2k$ of both, to account for the cost of \ocamli{f 1}.
Next, we face a par primitive, to execute \ocamli{left()} and \ocamli{right()}
in parallel.
We observe that the continuation of the program
is a call to \ocamli{f 5}, which has a cost of $k$.
Hence, we split the remaining $3k+1$
span credits into $k$ for the continuation (which we transfer with
\RULE{Transfer}) and $2k+1$ for the upcoming primitive parallel operation.
While we're at it, we also split the remaining $6k+1$ work credits
into~$k$ for the continuation (which we transfer with a standard \RULE{Frame} rule)
and $5k+1$ for the parallel operation.
We then apply \RULE{Par}, giving $2k+1$ span credits to the two sub-tasks.
We also split the $5k+1$ work credits into
$2k$ for \ocamli{left} and $3k+1$ for \ocamli{right}.
We first verify \texttt{left()},
which calls \ocamli{f 2} and consumes $2k$ work credits and $2k$ span credits (which we have, since we have $2k+1$ span credits available).
We then verify
\texttt{right()},
which executes a tick (consuming $1$ work credit and~$1$ span credit)
and calls in parallel \ocamli{f 3} and \ocamli{f 4}.
We again use \RULE{Par} and conclude this subgoal.
We finally go back to verify the continuation call to \ocamli{f 5}
and conclude.

Now that we have seen the key ideas behind \logic{}
we turn to a more formal presentation of the language
under consideration~(\Cref{sec:syntax_and_semantics})
and the proof rules~(\Cref{sec:logic}).

\section{Syntax and Semantics}
\label{sec:syntax_and_semantics}
We call the formal language we study \lang.
\lang is a call-by-value lambda calculus
with mutable state and parallelism.
We first present its syntax~(\cref{sec:syntax})
and then its semantics~(\cref{sec:semantics}).
\lang is similar to \heaplang, the language
that ships with Iris, except that it implements structured parallelism
instead of fork-based concurrency, and provides a $\etick$ primitive
for cost accounting.

\subsection{Syntax}
\label{sec:syntax}
\begin{figure}\centering\small
\newcommand{\commentary}[1]{ & \text{\small\it #1} \\}
\[
\begin{array}{l@{\quad}r@{\;\defineq\;}l}
\text{Values}\;\Values &\val & \vunit \mid \vbool \in \{\vtrue,\vfalse\} \mid \vint \in \mathbb{Z} \mid \loc \in \Loc \mid \vfun{f}{x}{\expr} \\

\text{Primitives} & \primitive  & + \mid - \mid \times \mid \div \mid\textsf{mod} \mid \symbeq \mid \mathord{<} \mid \mathord{\leq} \mid \mathord{>} \mid \mathord{\geq} \mid \symbor \mid \symband \\

\text{Expressions} &\expr&
\begin{array}[t]{@{}l@{\hspace{8mm}}l@{}}
\begin{array}[t]{@{}ll@{}}
\val,\wal
\commentary{value}
\var
\commentary{variable}
\elet{\var}\expr\expr
\commentary{sequencing}
\eif\expr\expr\expr
\commentary{conditional}
\efun{f}{x}\expr
\commentary{abstraction}
\eapp\expr{\expr}
\commentary{call}
\ecallprim\expr\expr
\commentary{primitive operation}
\end{array} &
\begin{array}[t]{@{}ll@{}}
\etick
\commentary{cost operation}
\ealloc\expr\expr
\commentary{array allocation}
\eload\expr\expr
\commentary{array load}
\estore\expr\expr\expr
\commentary{array store}
\elength\expr
\commentary{array length}
\epar\expr\expr
\commentary{parallelism}
\ecas\expr\expr\expr\expr
\commentary{compare-and-swap}
\end{array}
\end{array}\\
\text{Contexts} & \ectx &
\begin{array}[t]{@{}lllll}
\elet{\var}\khole\expr &\mid \eif\khole\expr\expr &\mid \ealloc\expr\khole &\mid \ealloc\khole\val &\mid \eload\expr\khole \\
\mid \eload\khole\val &\mid \estore\expr\expr\khole &\mid \estore\expr\khole\val &\mid \estore\khole\val\val &\mid \elength\khole \\
\mid \eapp{\expr}\khole &\mid \eapp\khole{\val} &\mid \ecallprim\expr\khole &\mid \ecallprim\khole\val \\
\mid \ecas\expr\expr\expr\khole &\mid \ecas\expr\expr\khole\val &\mid \ecas\expr\khole\val\val &\mid \ecas\khole\val\val\val
\end{array}
\end{array}\]
\captionlabel{Syntax of \lang}{fig:syntax}
\end{figure}

\cref{fig:syntax} presents the syntax of \lang.
A value~$\val \in \Values$ is either the unit value~$\vunit$, a Boolean~$\vbool \in \{\vtrue,\vfalse\}$,
an idealized integer~$\vint \in \mathbb{Z}$,
a location~$\loc$ from an infinite set of locations $\Loc$,
or a recursive function~$\vfun{f}{x}{\expr}$.
A computation in \lang
is described by an expression~$\expr$,
whose syntax is mostly standard.
Recursive functions are written~$\efun{f}{x}{\expr}$ and will evaluate
to their value counterpart.
We write~$\efunnonrec{x}{\expr} \eqdef \efun{\_}{x}{\expr}$ for non-recursive functions,
and define functions with multiple arguments as a chain of function constructors.
The $\etick$ operation is used for cost accounting.
Mutable state is available through arrays.
The expression $\ealloc{\expr_1}{\expr_2}$ allocates an array of size~$\expr_1$
initialized with whatever value~$\expr_2$ evaluates to in each cell.
For the sake of simplicity at the logical level,
arrays are initialized, but the cost of initialization
is not accounted for.
This is unrealistic for large arrays, where initialization
typically costs $\complex{N}$ work and $\complex{\log N}$ span
for an array of size $N$.
If a verified program relies on the fact that the array
is initialized, then cost should be adequately incurred
using $\etick$ operations placed before the allocation point.
Parallelism is available with a primitive $\epar{\expr_1}{\expr_2}$,
which evaluates the two expressions in parallel and returns
their results as an array of size 2.
\lang also has a primitive compare-and-swap instruction $\ecas{\expr_1}{\expr_2}{\expr_3}{\expr_4}$,
which targets an array entry and has 4 parameters:
the array location, the offset into the array, the old value, and the new~value.

An evaluation context~$\ectx$ describes an expression with a hole~$\khole$
and dictates the right-to-left evaluation order of \lang.

\subsection{Semantics}
\label{sec:semantics}

\begin{figure}\centering\small
\begin{mathpar}
\inferrule[HeadIfTrue]{}
{\headstep{\eif{\vtrue}{\expr_1}{\expr_2}}\store{\expr_1}\store}

\inferrule[HeadIfFalse]{}
{\headstep{\eif{\vfalse}{\expr_1}{\expr_2}}\store{\expr_2}\store}

\inferrule[HeadCallPrim]
{\purestep{\ecallprim{\val_1}{\val_2}}\val}
{\headstep{\ecallprim{\val_1}{\val_2}}{\store}{\val}{\store}}

\inferrule[HeadClosure]{}
{\headstep{\efun{f}{x}{\expr}}{\store}{\vfun{f}{x}{\expr}}{\store}}

\inferrule[HeadCall]{}
{\headstep{\ecall{(\vfun{f}{x}{\expr})}{\val}}{\store}{\subst{f}{\vfun{f}{x}{\expr}}{\subst{x}{\val}{\expr}}}{\store}}

\inferrule[HeadLetVal]{}
{\headstep{\elet{x}{\val}{\expr}}\store{\subst{x}\val{\expr}}\store}

\inferrule[HeadAlloc]
{0 < \vint \\ \loc \notin \dom\store}
{\headstep{\ealloc{\vint}{\val}}{\store}{\loc}{\blockupd\store{\loc}{\blockrepeat{\vint}{\val}}}}

\inferrule[HeadLength]
{\store(\loc) = \wals \\ \vint = \length\wals}
{\headstep{\elength{\loc}}{\store}{\vint}{\store}}

\inferrule[HeadLoad]
{\store(\loc) = \wals \\ 0 \leq \vint < \length\wals \\ \wals(i) = \val}
{\headstep{\eload{\loc}{\vint}}{\store}{\val}{\store}}

\inferrule[HeadStore]
{\store(\loc) = \wals \\ 0 \leq \vint < \length\wals}
{\headstep{\estore{\loc}{\vint}{\val}}\store\vunit{\blockupd{\store}{\loc}{\blockupd\wals{\vint}{\val}}}}

\inferrule[HeadCASSucc]
{ \store(\loc) = \wals \\
  0 \leq \vint < \length\wals \\
  \wals(\vint) = \val
}
{\headstep{\ecas\loc{\vint}\val{\val'}}\store\vtrue{\blockupd\store\loc{\blockupd\wals\vint{\val'}}}}

\inferrule[HeadCASFail]
{ \store(\loc) = \wals \\
  0 \leq \vint < \length\wals \\
  \wals(\vint) = \val_0 \\
  \val_0 \neq \val
}
{\headstep{\ecas\loc{\vint}\val{\val'}}\store\vfalse\store}
\end{mathpar}
\captionlabel{Head reduction relation}{fig:headsemantics}
\end{figure}

\paragraph{Head reduction relation}
\Cref{fig:headsemantics} presents the head reduction
relation~$\smash{\headstep{\expr}{\store}{\expr'}{\store'}}$,
describing a single step of expression~$\expr$ with store
$\store$ into expression~$\expr'$ and store~$\store'$.
A store is a map from locations to arrays,
which are modeled as lists of values.
We write $\emptyset$ for the empty store and $\store(\loc)$
for the array at location $\loc$ in~$\store$.
To insert or update a location~$\loc$ with array $\vals$ in store~$\store$,
we write $\blockupd{\store}{\loc}{\vals}$, and similarly write $\blockupd{\vals}{\vint}{\wal}$ to update
offset $\vint$ with value $\wal$ in array $\vals$.
The length of an array $\vals$ is written as $\length\vals$, and $\blockrepeat{\vint}{\val}$
represents an array of size $\vint$ initialized with value $\val$ in each cell.

The reduction rules are standard.
\RULE{HeadIfTrue} and \RULE{HeadIfFalse} handle conditionals.
\RULE{HeadCallPrim} evaluates primitive operations.
\RULE{HeadClosure} converts a function expression into a function value.
\RULE{HeadCall} performs function application.
\RULE{HeadLetVal} substitutes the bound value into the body.
\RULE{HeadAlloc} allocates an array initialized with the given value and returns its location, which is selected nondeterministically.
\RULE{HeadLoad} and \RULE{HeadStore} perform array loads and stores, respectively.
\RULE{HeadLength} returns the length of an array.
\RULE{HeadCASSucc} and \RULE{HeadCASFail} perform an atomic compare-and-swap.

\paragraph{Scheduler reduction relation}
\begin{figure}
\centering\small
\begin{mathpar}
\inferrule[SchedHead]
{\headstep\expr\store{\expr'}{\store'}}
{\parastep{\smallconfig\expr{\tleaf\vertex}{\phruple\store\tickmap\graph}}{\smallconfig{\expr'}{\tleaf\vertex}{\phruple{\store'}{\tickmap}\graph}}}

\inferrule[SchedTick]
{\tickmap' = \blockupd\tickmap\vertex{\tickmap(\vertex)+1}}
{\parastep{\smallconfig\etick{\tleaf\vertex}{\phruple\store\tickmap\graph}}
{\smallconfig{\vunit}{\tleaf\vertex}{\phruple{\store}{\tickmap'}\graph}}}

\inferrule[SchedFork]
{\vertex_1,\vertex_2 \notin \vertices\graph \\
  \tickmap' = \blockupd{\blockupd{\tickmap}{\vertex_1}{0}}{\vertex_2}{0} \\
  \graph' = \graph \cup \{\graphedge{\vertex_0}{\vertex_1},\graphedge{\vertex_0}{\vertex_2} \}
}
{\parastep
  {\smallconfig{\epar{\expr_1}{\expr_2}}{\tleaf{\vertex_0}}{\phruple\store\tickmap\graph}}
  {\smallconfig{\epar{\expr_1}{\expr_2}}{\littlenode{\vertex_1}{\vertex_2}}{\phruple{\store}{\tickmap'}{\graph'}}}}

\inferrule[SchedJoin]
  { \loc \notin \dom{\store} \\
    \vertex_3 \notin \dom{\tickmap}  \\
    \vertex_3 \notin \vertices\graph \\
   \graph' = \graph \cup \{\graphedge{\vertex_1}{\vertex_3},\graphedge{\vertex_2}{\vertex_3} \}
  }
  {\parastep
    {\smallconfig{\epar{\val_1}{\val_2}}{\littlenode{\vertex_1}{\vertex_2}}{\phruple\store\tickmap\graph}}
    {\smallconfig{\loc}{\tleaf{\vertex_3}}{\phruple{\blockupd\store\loc{[\val_1;\val_2]}}{\blockupd\tickmap{\vertex_3}{0}}{\graph'}}}}

\inferrule[StepSched]
{\parastep
  {\smallconfig\expr\tasktree{\phruple\store\tickmap\graph}}
  {\smallconfig{\expr'}{\tasktree'}{\phruple{\store'}{\tickmap'}{\graph'}}}}
{\step
  {\smallconfig\expr\tasktree{\thruple\store\tickmap\graph}}
  {\smallconfig{\expr'}{\tasktree'}{\thruple{\store'}{\tickmap'}{\graph'}}}}

\inferrule[StepBind]
{\step
  {\smallconfig\expr\tasktree\state}
  {\smallconfig{\expr'}{\tasktree'}{\state'}}}
{\step
  {\smallconfig{\efillctx\ectx\expr}\tasktree\state}
  {\smallconfig{\efillctx\ectx{\expr'}}{\tasktree'}{\state'}}}

\inferrule[StepParL]
{\step
  {\smallconfig{\expr_1}{\tasktree_1}\state}
  {\smallconfig{\expr_1'}{\tasktree_1'}{\state'}}}
{\step
  {\smallconfig{\epar{\expr_1}{\expr_2}}{\tpar{\tasktree_1}{\tasktree_2}}\state}
  {\smallconfig{\epar{\expr_1'}{\expr_2}}{\tpar{\tasktree_1'}{\tasktree_2}}{\state'}}}

\inferrule[StepParR]
{\step
  {\smallconfig{\expr_2}{\tasktree_2}\state}
  {\smallconfig{\expr_2'}{\tasktree_2'}{\state'}}}
{\step
  {\smallconfig{\epar{\expr_1}{\expr_2}}{\tpar{\tasktree_1}{\tasktree_2}}\state}
  {\smallconfig{\epar{\expr_1}{\expr_2'}}{\tpar{\tasktree_1}{\tasktree_2'}}{\state'}}}
\end{mathpar}
\captionlabel{Reduction under a context and parallelism}{fig:semantics}
\end{figure}

During execution, each task has an associated \emph{task identifier}.
To keep track of the identifier of each task
and whether it started executing or not,
we follow~\citet{westrick-arora-acar-22} and use a \emph{task tree}.
A task tree $\tasktree \eqdef \tleaf{\vertex} \mid \tpar{\tasktree_1}{\tasktree_2}$
is either a leaf labeled with a task identifier~$\vertex$,
or a parallel composition of two sub-trees $\tasktree_1$ and $\tasktree_2$.
Leaves represent individual tasks, while internal nodes represent
\texttt{par} expressions whose children are currently executing in parallel.
Vertices of task trees and their dependencies are recorded
in a \emph{computation graph}~$\graph$, modeled as a set of edges.
Weights of vertices (their number of executed $\etick$ operations) are
recorded by a \emph{tick map}~$\tickmap$.
The upper part of \Cref{fig:semantics} presents the scheduling relation
$\smash{\parastep{\smallconfig{\expr}{\tasktree}{\phruple{\store}{\tickmap}{\graph}}}{\smallconfig{\expr'}{\tasktree'}{\phruple{\store'}{\tickmap'}{\graph'}}}}$
reducing expression~$\expr$ with a task tree~$\tasktree$,
a store~$\store$, a tick map~$\tickmap$, and a computation graph~$\graph$.
\RULE{SchedHead} performs a head step.
\RULE{SchedTick} increments the tick count for the current task and returns the unit value.
\RULE{SchedFork}
converts a leaf executing a parallel primitive
into a node with two leaves.
In the computation graph, these two leaves form
two fresh vertices with edges from the original task and an initial cost of $0$.
\RULE{SchedJoin} joins a node with two leaves where both sides reached a value
into a new leaf, returning a location pointing to an array storing the two values.
In the computation graph, this leaf forms a fresh vertex
with edges from the two previous tasks.

Note that these joins are the only intended primitive for a thread to wait or synchronize with another thread without generating additional cost.
Because the language has shared mutable state, programs can also implement a form of busy-waiting between threads by spinning on a shared reference.
However, it is important to include a $\etick$ in any such loop, or else the resulting computation graph would have an unrealistic form of cost-free synchronization between threads.

\paragraph{Main reduction relation}
The lower part of \Cref{fig:semantics} presents the main reduction relation, written
$\smash{\step{\smallconfig{\expr}{\tasktree}{\state}}{\smallconfig{\expr'}{\tasktree'}{\state'}}}$,
where $\state = \thruple{\store}{\tickmap}{\graph}$ abbreviates the shared state.
\RULE{StepSched} lifts a scheduling step to the main reduction relation.
\RULE{StepBind} performs a step under an evaluation context.
\RULE{StepParL} and \RULE{StepParR} implement parallelism:
these two rules allow the main reduction relation to nondeterministically
step either the left or right side of an active parallel composition.
We write the reflexive-transitive closure
of the reduction relation as $\smash{\steprtc{\smallconfig{\expr}{\tasktree}{\state}}{\smallconfig{\expr'}{\tasktree'}{\state'}}}$.

\section{The Program Logic and its Soundness Theorem}
\label{sec:logic}
In this section,
we present the non-credit reasoning rules of \logic that were not previously shown~(\cref{sec:rules}),
and explain how we derive \RULE{Transfer}
from more primitive rules~(\cref{sec:innertransfer}).
We next present the formal statement of the soundness theorem of \logic~(\cref{sec:soundness_thm}).
We then turn to the proof of this theorem and its technical details:
we define the weakest precondition~(\cref{sec:semantic_model}),
and set up the state interpretation predicate relating physical state to logical state~(\cref{sec:interp}),
focusing in particular on our handling of credits~(\cref{sec:interp:credits}).

\subsection{Reasoning Rules of \logic}
\label{sec:rules}
\begin{figure}
\centering\small\morespacingaroundstar
\begin{mathpar}
\inferrule[Value]
{\post\,\vertex\,\val}
{\wpg\vertex\val\post}

\inferrule[Alloc]
{\pure{0 < \allocsize}}
{\wpgex{\vertex}{(\ealloc\allocsize\val)}{\_}{\loc}
  {\loc\mapsto{\blockrepeat\allocsize\val}}}

\inferrule[Load]
{\pure{0 \leq \ofs < \length\wals \land \wals(\ofs) = \val} \\
  \loc \mapsto_{\qp} \wals}
{\wpgex{\vertex}{(\eload{\loc}{\ofs})}{\_}{\val'}
  {\pure{\val'=\val} \star \loc \mapsto_{\qp}\wals}}

\inferrule[Store]
{\pure{0 \leq \ofs < \length\wals} \\ \loc \mapsto \wals}
{\wpgex{\vertex}{(\estore{\loc}{\ofs}{\val})}{\_}{\val'}
  {\pure{\val'=\vunit} \star \loc \mapsto \blockupd\wals\ofs\val}}

\inferrule[Length]
{\loc\mapsto_{\qp}\wals}
{\wpgex{\vertex}{(\elength\loc)}{\_}{\vint}
  {\pure{\vint = \length\wals} \star \loc\mapsto_{\qp}\wals}}

\inferrule[IfTrue]
{\wpg\vertex{\expr_1}\post}
{\wpg\vertex{(\eif\vtrue{\expr_1}{\expr_2})}\post}

\inferrule[IfFalse]
{\wpg\vertex{\expr_2}\post}
{\wpg\vertex{(\eif\vfalse{\expr_1}{\expr_2})}\post}

\inferrule[LetVal]
{\wpg{\vertex}{(\subst\var\val\expr)}{\post}}
{\wpg{\vertex}{(\elet\var\val\expr)}{\post}}

\inferrule[Bind]
{\wpgex{\vertex}{\expr}{\vertex'}{\val}{\wpg{\vertex'}{(\efillctx\ectx\val)}\post}}
{\wpg{\vertex}{(\efillctx\ectx\expr)}{\post}}

\inferrule[CallPrim]
{\pure{\purestep{\ecallprim{\val_1}{\val_2}}{\val}}}
{\wpgex\vertex{(\ecallprim{\val_1}{\val_2})}{\_}{\val'}{\pure{\val' = \val}}}

\inferrule[Call]
{\pure{\val = \vfun{\funcname}{\argname}{\body}} \\
\later \wpg{\vertex}{(\subst\funcname\val{\subst\argname{\val'}\body})}{\post}
}
{\wpg{\vertex}{(\eapp\val{\val'})}\post}

\inferrule[CASSucc]
{\pure{0 \leq \ofs < \length\wals \land \wals(\ofs) = \val_0 \land \val_0 = \val} \\
  \loc \mapsto\wals}
{\wpgex\vertex{(\ecas\loc\ofs\val{\val'})}{\_}{\vbool}
  {\pure{\vbool=\vtrue} \star \loc \mapsto \blockupd\wals\ofs{\val'}}}

\inferrule[CASFail]
{\pure{0 \leq \ofs < \length\wals \land \wals(\ofs) = \val_0 \land \val_0 \neq \val} \\
  \loc \mapsto_{\qp}\wals}
{\wpgex\vertex{(\ecas\loc\ofs\val{\val'})}{\_}{\vbool}
  {\pure{\vbool=\vfalse} \star \loc \mapsto_{\qp} \wals}}

\inferrule[Closure]{}
{\wpgex{\vertex}{(\efun\funcname\argname\body)}{\_}{\val}
  {\pure{\val=\vfun{\funcname}{\argname}{\body}}}}

\inferrule[Atomic]
{\pure{\mathsf{atomic}\,\expr} \\
  \wpgex{\vertex}{\expr}{\vertex'}{\val}{\pure{\vertex'=\vertex} \wand \post\,\vertex\,\val}}
{\wpg{\vertex}{\expr}{\post}}

\inferrule[Frame]
{\pre_0 \\ \wpgex\vertex\expr{\vertex'}{\val}{\pre_1}}
{\wpgex\vertex\expr{\vertex'}{\val}{\pre_0 \star \pre_1}}
\end{mathpar}
\captionlabel{Reasoning rules for base constructs, except for the tick and the parallel primitives}{fig:rules}
\end{figure}

\Cref{fig:rules} presents
reasoning rules for basic constructs,
except for the tick and the parallel primitive that were explained previously~(see \cref{sec:key:credits}).
Apart from the task identifier, they are mostly standard.
\logic makes use of
fractional~\citep{boyland-fractions-03,bornat-permission-accounting-05}
points-to assertions of the form~$\loc
\mapsto_{\qp} \wals$, where~$\qp$ denotes a positive fraction
less than or equal to 1.
When~$\qp = 1$ we write~$\loc \mapsto \wals$.
\RULE{Value} ensures that when the execution reaches a value, the postcondition is satisfied.
\RULE{Alloc}, \RULE{Load}, \RULE{Store}, and \RULE{Length} are the usual array operations, with \RULE{Load} and \RULE{Length} requiring only fractional ownership.
\RULE{IfTrue} and \RULE{IfFalse} handle conditionals.
\RULE{LetVal} substitutes a value in a let binding.
\RULE{Bind} is the sequencing rule for evaluation contexts.
\RULE{CallPrim} handles primitive operations, and \RULE{Call} handles function application, with a later modality $\later$~\citep{iris}.
\RULE{Closure} creates function values.
\RULE{CASSucc} and \RULE{CASFail} describe a successful and a failed CAS, respectively.

For conciseness, the above rules do not mention
that sequential operations preserve the identifier of the current task.
\RULE{Atomic} addresses this: for
any atomic operation (\ie, one that reduces to a value in exactly one step),
the end task is the same as the current task.

The WP of \logic
also satisfies other standard rules of separation logic,
such as the~\RULE{Frame} rule.

\subsection{Inside the Transfer Rule}
\label{sec:innertransfer}
\begin{figure}\centering\small\morespacingaroundstar\morespacingaroundwedge
\begin{mathpar}
\inferrule[GenerateTransferable]
{\wpgex{\vertex}{\expr}{\vertex'}{\val}{\cantransfer{\vertex}{\vertex'} \wand \post\,\vertex'\,\val}}
{\wpg{\vertex}{\expr}{\post}}

\inferrule[PrimitiveTransfer]
{\cantransfer{\vertex}{\vertex'} \\ \spanc{\vertex}{n} \\\\ (\spanc{\vertex'}{n} \wand \wpg{\vertex'}{\expr}{\post})}
{\wpg{\vertex'}{\expr}{\post}}

\inferrule[End]
{\wpgex\vertex\expr{\vertex'}{\val}{\wpg{\vertex'}\val\post}}
{\wpg\vertex\expr\post}

\inferrule[TF-Refl]{}
{\cantransfer{\vertex}{\vertex}}

\inferrule[TF-Trans]
{\cantransfer{\vertex}{\vertex''} \\
\cantransfer{\vertex''}{\vertex'}}
{\cantransfer{\vertex}{\vertex'}}

\inferrule[TF-Diamond]
{\cantransfer{\vertex_{l}}{\vertex'_{l}} \\
\cantransfer{\vertex_{r}}{\vertex'_{r}} \\\\
\isfork{\vertex}{\vertex_{l}}{\vertex_{r}} \\
\isjoin{\vertex'_{l}}{\vertex'_{r}}{\vertex'}}
{\cantransfer{\vertex}{\vertex'}}
\end{mathpar}
\captionlabel{Reasoning principles for \cantransfername{}}{fig:transferproof}
\end{figure}

The \RULE{Transfer} rule is derived from
more primitive and slightly more flexible reasoning rules,
presented in \Cref{fig:transferproof}.
The rules make use of the persistent (hence duplicable)
assertion $\cantransfer{\vertex}{\vertex'}$,
which asserts that span credits can be transferred from $\vertex$ to $\vertex'$.
The \cantransfername assertion is reflexive (\RULE{TF-Refl})
and transitive (\RULE{TF-Trans}).

\RULE{GenerateTransferable}
allows the user to generate permission
to transfer span credits from
the current task $\vertex$ to the end task $\vertex'$,
by producing an assertion $\cantransfer{\vertex}{\vertex'}$
in the postcondition.
\RULE{PrimitiveTransfer}
actively transfers credits from one task to another:
it requires a witness $\cantransfer{\vertex}{\vertex'}$,
as well as span credits $\spanc{\vertex}{n}$
while facing a weakest precondition for $\vertex'$,
consumes these two previous assertions
and gives back $\spanc{\vertex'}{n}$ span credits for pursuing the proof.
Together, \RULE{GenerateTransferable} and \RULE{PrimitiveTransfer}
are more flexible than \RULE{Transfer}.
They allow for postponing the transfer
of span credits to a later point in the proof,
rather than manually applying \RULE{Transfer}
before verifying expressions involving a parallel primitive.
\RULE{End} is a technical rule
that, read from bottom to top, creates at will
a trivial weakest precondition on a value in the postcondition.
\RULE{End} is useful since \RULE{PrimitiveTransfer}
requires the goal to be a weakest precondition.

Equipped with these rules, we derive \RULE{Transfer} as follows:
first we apply \RULE{End} to generate a trivial WP in the postcondition,
then we apply \RULE{GenerateTransferable} to generate an assertion $\cantransfer{\vertex}{\vertex'}$,
next we apply \RULE{Frame} to move span credits in the postcondition,
and then move the reasoning to the postcondition,
where we apply \RULE{PrimitiveTransfer} and \RULE{Value} to conclude.

The lower part of \Cref{fig:transferproof} presents
the definition of the assertion $\cantransfer{\vertex}{\vertex'}$, which is defined
inductively as a least fixed point of the rules shown in that figure.
\RULE{TF-Refl} and \RULE{TF-Trans} state that the relation is reflexive and transitive, respectively.
The most complex rule is \RULE{TF-Diamond},
which in turn uses more primitive assertions $\isfork{\vertex}{\vertex_l}{\vertex_r}$
and $\isjoin{\vertex'_l}{\vertex'_r}{\vertex'}$, which describe the corresponding properties
of the structure of the computation graph.
In particular, this rule captures the case of a ``diamond''
in the computation graph:
the task $\vertex$ forks into two tasks~$\vertex_l$ and~$\vertex_r$,
which execute, possibly forking and joining,
ultimately reaching tasks~$\vertex'_l$ and~$\vertex'_r$ that join into $\vertex'$.
Although ``end users'' of the logic will typically use the simple transfer rule we saw in \Cref{sec:key_ideas},
deriving this rule out of these more primitive concepts simplifies the soundness proof that we carry out later.

\subsection{Soundness Theorem}
\label{sec:soundness_thm}

To state the soundness theorem of \logic{},
we first explain
how to express bounds on the work and span of programs,
and then define
what it means for an expression to be safe (\ie, to not crash).

\paragraph{Formal bounds on work and span}
\begin{figure}
\centering\small\morespacingaroundstar
\begin{mathpar}
\inferrule[WorkBound]
{\sumall\tickmap \leq n}
{\workbound\tickmap{n}}

\inferrule[PathNil]
{\tickmap(\vertex) = n}
{\ispath\graph\tickmap\vertex{n}\vertex}

\inferrule[PathCons]
{(\vertex_1,\vertex_2) \in \graph \\
\tickmap(\vertex_1) = n \\\\
\ispath\graph\tickmap{\vertex_2}{n'}{\vertex_3}}
{\ispath\graph\tickmap{\vertex_1}{n+n'}{\vertex_3}}

\inferrule[SpanBound]
{\forall \vertex_1\,\vertex_2\,n'.\quad\ispath\graph\tickmap{\vertex_1}{n'}{\vertex_2} \implies n' \leq n}
{\spanbound\graph\tickmap{n}}
\end{mathpar}
\captionlabel{Definitions of the \workboundname and \spanboundname predicates}{fig:workspan}
\end{figure}

\Cref{fig:workspan} presents the \workboundname and \spanboundname predicates.
The property $\workbound{\tickmap}{n}$ holds when the work---that is, the sum of all ticks recorded in the tick map $\tickmap$---is
at most $n$~(\RULE{WorkBound}).

Bounding the span requires bounding the cost along any execution path in the task graph.
To this end, we first define an auxiliary predicate
$\ispath{\graph}{\tickmap}{\vertex_1}{n}{\vertex_2}$,
which asserts that there exists a path from $\vertex_1$ to $\vertex_2$ in $\graph$
whose total tick count (summing $\tickmap(\vertex)$ for each vertex $\vertex$ on the path) is $n$.
\RULE{PathNil} handles the base case where the path consists of a single vertex
($\vertex_1 = \vertex_2$), and \RULE{PathCons}
extends a path by prepending an edge.
The property $\spanbound{\graph}{\tickmap}{n}$
holds when every path in the graph has a total tick count of at most $n$~(\RULE{SpanBound}).

\paragraph{Safety of an expression}
\begin{figure}
\centering\small
\begin{mathpar}
\inferrule[RedSched]
  {\parastep
  {\smallconfig\expr\tasktree\state}
  {\smallconfig{\expr'}{\tasktree'}{\state'}}}
  {\reducible\expr\tasktree\state}

\inferrule[RedCtx]
  {\reducible\expr\tasktree\state}
  {\reducible{(\efillctx\ectx\expr)}\tasktree\state}

\inferrule[RedPar]
  {(\expr_1 \notin \Values \;\lor\; \expr_2 \notin \Values ) \\\\
  (\expr_1 \notin \Values \implies \reducible{\expr_1}{\tasktree_1}\state) \\
  (\expr_2 \notin \Values \implies \reducible{\expr_2}{\tasktree_2}\state)}
  {\reducible{(\epar{\expr_1}{\expr_2})}{(\tpar{\tasktree_1}{\tasktree_2})}\state}

\inferrule[Safe]
{ (\expr \in \Values \wedge \tasktree \in \vertexs) \quad\lor\quad \reducible\expr\tasktree\state}
{ \safe\expr\tasktree\state }
\end{mathpar}
\captionlabel{Reducibility and safety of a configuration}{fig:reducible}
\end{figure}

\Cref{fig:reducible} first presents
the notion of reducibility for \lang.
Intuitively, a configuration satisfies all-task reducibility,
written $\reducible{\expr}{\tasktree}{\state}$
if \emph{every} task of expression $\expr$ described by the task tree $\tasktree$
can take a step with state $\state$.
\RULE{RedSched} asserts that any expression that can take a scheduling step is
all-task reducible, since there is a single task and it can take a step.
\RULE{RedCtx} accounts for evaluation contexts.
\RULE{RedPar} asserts that, when facing a parallel primitive $\epar{\expr_1}{\expr_2}$ with two subtrees
$\tasktree_1$ and $\tasktree_2$, then if $\expr_1$ is not a value,
then it should be all-task reducible with~$\tasktree_1$,
if $\expr_2$ is not a value, then it should be all-task reducible with $\tasktree_2$,
and at least one of $\expr_1$ and $\expr_2$ has to not be a value (otherwise, a join is possible).
\Cref{fig:reducible} then presents the notion of safety.
\RULE{Safe} asserts that an expression $\expr$ is safe with a task tree $\tasktree$
and a state $\state$,
written $\safe{\expr}{\tasktree}{\state}$,
if either the expression is a value and the task tree a single leaf,
or the configuration is all-task reducible.

\paragraph{Soundness statement}
We can then state the soundness of \logic, which asserts that,
if the user verified the program $\expr$ with $w$ initial work credits and $s$ initial span credits,
then
(1) the program is always safe,
(2) the work is bounded by $w$, and
(3) the span is bounded by $s$.

\begin{theorem}[Soundness of \logic]
\label{thm:soundness}
If $\workc{w} \star \spanc{\vertex_0}{s} \vdash \wpgex{\vertex_0}{\expr}{\_}{\_}{\iTrue}$ holds,
and if $\steprtc{\smallconfig\expr{\vertex_0}{\thruple{\emptyset}{\singlemap{\vertex_0}{0}}{\emptyset}}}{\smallconfig{\expr'}\tasktree{\thruple\store\tickmap\graph}}$
then:
\begin{enumerate}
\item $\safe{\expr'}{\tasktree}{\thruple\store\tickmap\graph}$ holds,
\item $\workbound{\tickmap}{w}$ holds, and
\item $\spanbound{\graph}{\tickmap}{s}$ holds too.
\end{enumerate}
\end{theorem}

As mentioned earlier, the relation $\steprtcname$ is non-deterministic, and so there are multiple possible tick maps $\tickmap$ and computation graphs $\graph$ that can arise from execution.
However, the bounds in the theorem above apply to \emph{any} such $\tickmap$ and $\graph$ obtained from executing $e$, and so they upper bound the work and span across all possible interleavings of steps.

The proof of this theorem follows the standard Iris recipe, as we explain in the next section.

\section{Proof of the Soundness Theorem}

This section explains the soundness theorem.
We start with a high-level overview of the structure~(\cref{sec:proof_overview}).
Then we define the weakest precondition~(\cref{sec:semantic_model})
and present the
properties of the key \emph{state interpretation predicate}~(\cref{sec:interp}),
which gives meaning to work and span credits and relates them
to the concrete work and span values~(\cref{sec:interp:credits}).

\subsection{Proof Overview}
\label{sec:proof_overview}

The proof of \Cref{thm:soundness} follows the standard Iris progress-and-preservation recipe~\citep[\S6.4]{iris},
which is made possible through the definition of the weakest precondition~(\cref{sec:semantic_model}).
Indeed, this definition
ensures that the state interpretation predicate~(\cref{sec:interp})
is maintained as an invariant between steps of the execution.
In particular, this predicate relates the physical state (the store, tick map, and computation graph)
to ghost resources (points-to assertions, work and span credits, \dots).
The WP guarantees that every configuration reachable by execution
will satisfy the state interpretation.
Using the properties of the state interpretation,
we can then prove the three properties of \Cref{thm:soundness}:
(1)~safety,~(2) the work bound, and (3)~the span bound,
presented in increasing order of proof difficulty.

\paragraph{Safety}
To verify safety, we extend the Iris WP
to account for the task tree~(\cref{sec:semantic_model}).

\paragraph{Work bound} We bound the work following the same technique
as for standard \emph{time credits}~\citep{atkey-11,mevel-jourdan-pottier-19}.
Each application of \RULE{Tick} consumes one work credit supplied by the user,
which the state interpretation stores away as a witness
that the user indeed paid for that unit of work.
Since the user starts with $w$ credits and credits are never created out of thin air,
the state interpretation may store at most~$w$ credits, and thus the work is bounded by $w$.

\paragraph{Span bound}
Span credits are annotated
with a task identifier,
and are tracked using one counter per task,
representing the amount of span credits in circulation for this task.
In order to bound the total span,
the state interpretation stores span credits in two places.
First, similarly to work credits,
the state interpretation stores one span credit
for the task~$\vertex$
upon each application of \RULE{Tick} for $\vertex$.
This credit forms a witness that the user
paid.
Second, the state interpretation stores span credits
expressing the relationship between
the span of a task and the span of its children.
More precisely, upon a \RULE{Par} that forks
tasks~$\vertex_1$ and~$\vertex_2$ from~$\vertex$,
the user gives up $n$ span credits of $\vertex$,
and obtains $n$ span credits
of $\vertex_1$ and $n$ span credits of $\vertex_2$.
The initial $n$ credits of $\vertex$ are stored inside the state interpretation,
as witness of the number of span credits available to $\vertex_1$ and $\vertex_2$.
Using these two invariants over credits,
we can prove that the number of span credits
available to each task bounds the heaviest path
starting from this task~(\cref{sec:interp:credits}).

The delicate point is how span credits flow
when a task~$\vertex$ transfers them
to another task~$\vertex'$, via \RULE{PrimitiveTransfer}.
Indeed, the second invariant described above seems to fix
forever the amount of span credits available for each task.
Our idea is to generate fresh credits in a \emph{cascade}:
starting from $\vertex$, we walk along every task on the paths leading to $\vertex'$,
and at each intermediate task we generate and deposit into
the state interpretation enough credits to justify the increase of its successors' credits,
thereby preserving the invariant.
This argument is sound in particular because the destination $\vertex'$
is the ``continuation'' of $\vertex$:
every path reaching $\vertex'$ must go through $\vertex$.
This property is captured by the assertion $\cantransfer{\vertex}{\vertex'}$
in the precondition of \RULE{PrimitiveTransfer}.

\subsection{Definition of the Weakest Precondition}
\label{sec:semantic_model}

Our definition of the WP and of assertions makes use of
ghost state,
relating the physical state of an execution to the
logical state of the proof.
The ghost state is updated with a \emph{ghost update},
written $\pvs$.
The assertion $\pvs \pre$ means
that $\pre$ holds after updating the ghost state.
As usual, we write $\pre \vs \pre'$
to mean $\square (\pre \wand \pvs \pre')$,
where $\square$ is the persistence modality.
Formally, ghost updates and our WP are parameterized by \emph{masks},
a standard syntactical device that controls which invariants
the user is allowed to open~\citep{iris}.
For brevity, we omit masks and refer the reader
to our formalization~\citep{mechanization}.

\begin{figure}\centering\small
\let\oldstar\star
\renewcommand{\star}{\,\oldstar\,}
\begin{align*}
&\kern0.74em\wpg{\vertex}{\expr}{\post} \;\eqdef\; \wpgen{\vertex}{\expr}{\post}\\
&\wpgen{\tasktree}{\expr}{\post} \;\eqdef\;
\forall\state.\; \stateinterp{\state}{\tasktree}{\expr} \vs\\
&\qquad\big(\, \pure{\expr \in \Values} \star
  \pvs\; \exists \vertex.\; \pure{\tasktree=\vertex} \star \stateinterp{\state}{\tasktree}{\expr} \star \post\;\vertex\;\expr
\,\big)\\
&\quad\lor\big(\,\pure{\expr \notin \Values} \star
  \pure{\reducible\expr\tasktree\state}\;\star
 \forall \state'\,\tasktree'\,\expr'.\; \pure{\smash{\step
  {\smallconfig\expr{\tasktree}{\state}}
  {\smallconfig{\expr'}{\tasktree'}{\state'}}}}\\
&\qquad\qquad \vs\;\later\;\pvs\;
  \stateinterp{\state'}{\tasktree'}{\expr'} \star \wpgen{\tasktree'}{\expr'}{\post}
\,\big)
\end{align*}
\captionlabel{Definition of the weakest precondition modalities}{fig:wp}
\end{figure}

The structure of the definition of our WP roughly follows the
one by \citet{dislog}.
The first line of \Cref{fig:wp} presents the definition of our WP.
First, we define $\wpg{\vertex}{\expr}{\post}$
in terms of a more general assertion,
written $\wpgen{\tasktree}{\expr}{\post}$,
parameterized not by a single task identifier~$\vertex$
but rather a whole task tree~$\tasktree$.
The remainder of \Cref{fig:wp} defines this general WP.
This definition makes use of
a state interpretation predicate $\stateinterp{\state}{\tasktree}{\expr}$,
which relates the physical state~$\state$, the task tree $\tasktree$ and the expression $\expr$
to the logical state;
we define it formally in \Cref{sec:interp}.
The definition quantifies over the state $\state$---recall that
the state encompasses the store, the tick map, and the computation graph~(\cref{sec:semantics})---%
and assumes ownership of the state
interpretation~$\stateinterp{\state}{\tasktree}{\expr}$.
Then, after a ghost update, two cases arise.
If $\expr$ is a value,
the task tree $\tasktree$ must be a single task $\vertex$,
the state interpretation must hold unchanged,
and the postcondition $\post\;\vertex\;\expr$ must hold too.
If $\expr$ is not a value,
it must be all-task reducible with $\tasktree$ and $\state$,
and for every possible step
to expression $\expr'$, task tree $\tasktree'$, and state~$\state'$,
then after several ghost updates, both
the state interpretation $\stateinterp{\state'}{\tasktree'}{\expr'}$
and the recursive call to the WP for $\expr'$ with $\tasktree'$ must hold.
The recursive call is guarded by a \emph{later} modality to ensure this recursive definition has a fixed point.

Because the general WP is indexed by an arbitrary task tree $T$, it can represent computations happening in parallel.
For example, the assertion $\wpgen{(\tpar{\tasktree_1}{\tasktree_2})}{(\epar{\expr_1}{\expr_2})}{\post}$
verifies that both sub-expressions $\expr_1$ and $\expr_2$
with their respective subtrees $\tasktree_1$ and $\tasktree_2$
can make progress in parallel.
This generality is used when deriving versions of \RULE{Par} or \RULE{Bind}.

In order to prove \Cref{thm:soundness},
we follow a \emph{progress and preservation} approach.
We prove three lemmas.
First, that the state interpretation
holds for the initial state, and the initial
amount of work and span credits can be given to the user.
Second, if the WP holds for some expression
and the state interpretation holds for some state,
then one step of the expression with this state
preserves the state interpretation and the WP.
Third, if the WP holds for some expression
and the state interpretation holds for some state,
then the work and span bounds hold,
and the expression is safe.

\subsection{State Interpretation}
\label{sec:interp}
\begin{figure}\centering\small
\morespacingaroundstar
\begin{align*}
\stateinterp{(\store,\tickmap,\graph)}{\tasktree}{\expr} \;\eqdef\;&
 \interpres{\store}{\tickmap}{\graph} \star \pure{\pureinv{\graph}{(\dom{\tickmap})}{\tasktree}{\expr}}
\\
\interpres{\store}{\tickmap}{\graph} \;\eqdef\;&
\heapinterp{\store} \star{} \interpwork{(\sumall{\tickmap})} \star \interpspan{\graph}{\tickmap} \star{} \tracktransfer{\graph}
\end{align*}
\captionlabel{Definition of the state interpretation predicate}{fig:interp}
\end{figure}

The state interpretation predicate $\stateinterp{S}{T}{\expr}$
relates the physical state $S=(\store,\tickmap,\graph)$ of an execution
of the program $\expr$ with task tree $\tasktree$
to the logical state of the proof.
This predicate holds at every execution step,
and its definition appears in \Cref{fig:interp}.
The state interpretation is composed of two parts:
one ``resourceful'' assertion $\interpresname$,
defining the ghost state, and
one ``pure'' part~$\pureinvname$,
asserting well-formedness properties of the different components.

The assertion $\interpres{\store}{\tickmap}{\graph}$
asserts ownership of~$\heapinterp\store$,
which gives meaning to points-to assertions;
this assertion is standard~\citep{genheap}.
The state interpretation next asserts ownership
of $\interpwork{(\sumall{\tickmap})}$,
which gives meaning to work credits---recall
that $\sumall{\tickmap}$ sums all ticks recorded in $\tickmap$---%
and of $\interpspan{\graph}{\tickmap}$, which gives meaning to span credits.
Both assertions are described in \Cref{sec:interp:credits}.
The state interpretation then asserts ownership
of $\tracktransfer{\graph}$.
This technical assertion
gives meaning to the assertions
$\isfork{\vertex}{\vertex_1}{\vertex_2}$
and $\isjoin{\vertex_1}{\vertex_2}{\vertex}$
that we saw in \Cref{sec:innertransfer}.

\begin{figure}\centering\small
\morespacingaroundwedge
\begin{align*}
\pureinv{\graph}{d}{\tasktree}{\expr}
  \;\triangleq\;\,&
   \hasnoloop{\graph} \wedge{} \vertices{\graph} \subseteq d
  \wedge{}\\
  &\leaves{\tasktree} \subseteq d
  \wedge \disjoint{\leaves{\tasktree}}{\sources{\graph}}
  \wedge \comptree{\tasktree}{\expr}
\end{align*}
\begin{mathpar}
\inferrule[CT-Leaf]{}
  {\comptree{\tleaf{\vertex}}{\expr}}

\inferrule[CT-Bind]
  {\comptree{\tasktree}{\expr}}
  {\comptree{\tasktree}{(\efillctx{\ectx}{\expr})}}

\inferrule[CT-Par]
  {\comptree{\tasktree_1}{\expr_1} \\
   \comptree{\tasktree_2}{\expr_2} \\\\
   \disjoint{\leaves{\tasktree_1}}{\leaves{\tasktree_2}}}
  {\comptree{(\tpar{\tasktree_1}{\tasktree_2})}{(\epar{\expr_1}{\expr_2})}}
\end{mathpar}
\captionlabel{Definition of pure invariants}{fig:pureinv}
\end{figure}

\Cref{fig:pureinv} presents the pure invariant
$\pureinv{\graph}{d}{\tasktree}{\expr}$,
with the argument $d$ naming the domain of the tick map $\tickmap$.
This invariant requires that the computation graph $\graph$
has no loop, that both its vertices and the leaves of the task tree $\tasktree$ are in $d$,
and that these leaves are not \emph{sources} in $\graph$---that is, they have no successors.
Moreover, the invariant requires
that $\tasktree$ is compatible with the expression $\expr$,
a fact denoted by $\comptree{\tasktree}{\expr}$.
This property is inductively defined in the lower part of
\Cref{fig:pureinv},
and intuitively asserts that there are
at least as many parallel expressions as nodes in the task tree,
and that the leaves of the tree are disjoint.
Indeed, \RULE{CT-Leaf} asserts that a leaf is compatible with any expression.
\RULE{CT-Bind} asserts that
if a task tree is compatible with an expression,
then it is also compatible with this expression within any evaluation context.
\RULE{CT-Par} asserts that
a node with two subtrees $\tasktree_1$ and $\tasktree_2$
is compatible with the parallel composition
of two expressions $\expr_1$ and $\expr_2$
if~$\tasktree_1$ is compatible with $\expr_1$,
$\tasktree_2$ is compatible with $\expr_2$,
and the leaves of $\tasktree_1$ and $\tasktree_2$ are disjoint.

\subsection{Interpretation of Credits}
\label{sec:interp:credits}
\newcommand{\TirNameStyleP}[1]{(\TirNameStyle{#1})}
\begin{figure}\centering\small\morespacingaroundstar
\[\begin{array}{cc}
\pvs\, \authwork{w_0} \star \workc{w_0} & \TirNameStyleP{WorkAlloc}\\
\authwork{w_0} \star \workc{w} \wand \pure{ w \leq w_0 } & \TirNameStyleP{WorkValid}
\end{array}\]
\begin{align*}
\interpwork{w} \;\eqdef\;& \workc{w} \star \exists w_0.\,\authwork{w_0} \star \initwork{w_0}
\end{align*}
\vspace{-1em}
\captionlabel{Ghost state for work credits}{fig:interp:work}
\vspace{0.5em}
\[\begin{array}{cc}
\pvs\, \authspan{\singlemap{\vertex_0}{s_0}} \star \spanc{\vertex_0}{s_0} & \TirNameStyleP{SpanAlloc}\\
\authspan\spaninitial \star \spanc{\vertex}{s} \wand \pure{ s \leq \spaninitial(\vertex) } & \TirNameStyleP{SpanValid}\\
\authspan\spaninitial \vs \authspan{(\updatecell\spaninitial{\vertex}{n})} \star \spanc{\vertex}{n} & \TirNameStyleP{SpanUpdate}
\end{array}\]
\begin{align*}
\interpspan{\graph}{\tickmap} \;\eqdef\;&
\big( \bigast{(\vertex,n) \in \tickmap}{\spanc{\vertex}{n}} \big) \star{}\\
&\exists \spaninitial.\;
\pure{\dom{\tickmap} = \dom{\spaninitial}} \star \authspan{\spaninitial} \star{} \initspan{\graph}{\spaninitial} \star \boundedsrc{\graph}{\spaninitial}
\\
\boundedsrc{\graph}{\tickmap} \;\eqdef\;&
\bigast{\vertex \in \sources{\graph}}{\exists n.\; \spanc{\vertex}{n} \star \pure{\forall \vertex'.\;
\graphedge{\vertex}{\vertex'} \in \graph \implies \tickmap(\vertex') \leq n}}
\end{align*}
\vspace{-1em}
\captionlabel{Ghost state for span credits}{fig:interp:span}
\end{figure}

In this section, we present the interpretation of work and span credits.
While in this description we axiomatize the ghost state for simplicity,
the seasoned Iris reader will recognize standard ghost state:
work credits are implemented with the camera $\authm(\nat)$
and span credits with the camera $\authm(\mapm(\taskids,\nat))$.

\paragraph{Work credits}
The upper part of \Cref{fig:interp:work}
presents the internal reasoning rules for work credits.
\RULE{WorkAlloc} allocates the initial amount of
work credits $w_0$,
producing both $\authwork{w_0}$ and $\workc{w_0}$
after a ghost update.
The assertion $\authwork{w_0}$
keeps track of the total amount of work credits in circulation:
\RULE{WorkValid} asserts that if $\authwork{w_0}$ and $\workc{w}$ hold,
then $w$ is at most~$w_0$.

The lower part of \Cref{fig:interp:work} defines
$\interpwork{w}$, where $w$ represents the work already done.
It asserts ownership of $\workc{w}$
and of $\authwork{w_0}$, for some existentially
quantified initial amount of work credits $w_0$.
The additional assertion $\initwork{w_0}$ is a proof artifact,
a persistent assertion that is used internally to remember
that the initial number of credits never changes.

Intuitively, each time \RULE{Tick} is applied in the proof,
the one work credit given by the user of the proof rule is stored in the assertion $\interpwork{w}$,
as a witness that the user indeed paid.
Hence, at any point during the execution,
one can use \RULE{WorkValid} to conclude that the work done so far is at most
the initial amount of work credits $w_0$.

\paragraph{Span credits}
The upper part of \Cref{fig:interp:span}
presents the reasoning rules for span credits.
\RULE{SpanAlloc} allocates
the initial amount of span credits $s_0$ for the initial task $\vertex_0$,
producing $\authspan{\singlemap{\vertex_0}{s_0}}$
and $\spanc{\vertex_0}{s_0}$ after a ghost update.
The assertion $\authspan\spaninitial$,
where $\spaninitial$ is a map from task identifiers to natural numbers,
keeps track of the total amount of span credits in circulation for each task.
\RULE{SpanValid} asserts that if $\authspan\spaninitial$ and $\spanc{\vertex}{s}$ hold,
then $s$ is at most $\spaninitial(\vertex)$.
As we explain in the next paragraph,
the total amount of span credits
available for a task may vary during the proof because of \RULE{PrimitiveTransfer},
except for the initial task.
To allow for that pattern, \RULE{SpanUpdate}
allows for updating the total amount
of span credits for a task $\vertex$ and generating said credits.
The function $\updatecell{\tickmap}{\vertex}{n}$
updates the map $\tickmap$ by setting the value of $\vertex$ to $n$ if $\vertex$ was not in $\tickmap$
or to $\tickmap(\vertex) + n$ otherwise.

The lower part of \Cref{fig:interp:span} defines
$\interpspan{\graph}{\tickmap}$,
where $\graph$ is the computation graph and $\tickmap$ the tick map.
It asserts ownership
of $\spanc{\vertex}{n}$ for each identifier $\vertex$ in the tick map,
where $n$ is the number of ticks recorded for $\vertex$ in $\tickmap$.
Second, it asserts the existence of a map
$\spaninitial$ from task identifiers to natural numbers,
with the same domain as $\tickmap$,
representing the
amount of span credits in circulation for each task,
as witnessed by $\authspan{\spaninitial}$.
The assertion $\initspan{\graph}{\spaninitial}$
is a proof artifact used internally to remember that
the amount of span credits for the initial task
(the only task without incoming edges in the computation graph)
does not change during the proof.
The assertion $\boundedsrc{\graph}{\spaninitial}$
captures the core of the interpretation of span credits:
for every source $\vertex$ of $\graph$
(that is, a vertex with at least one outgoing edge),
there exists some amount of span credit $n$ of~$\vertex$
such that, for any successor $\vertex'$ of~$\vertex$ in $\graph$,
$n$ bounds the amount of span credits for $\vertex'$.

Intuitively, each time \RULE{Tick} is applied in the proof,
the one span credit given by the user is stored in the assertion $\interpspan{\graph}{\tickmap}$,
as a witness that the user indeed paid.
Moreover, each time \RULE{Par} or \RULE{Transfer} is applied,
the span credits of the parent task given by the user
are stored in $\boundedsrc{\graph}{\spaninitial}$,
as a witness that
the span of the parent bounds
the span of all of its children.
Hence, at any point during the execution,
one can use \RULE{SpanValid} to conclude that
the initial amount of span credits
of the initial task bounds the weight of the heaviest
path starting from it.

The reader might wonder what happens
during the join of two tasks $\vertex_1$ and $\vertex_2$
into a continuation $\vertex'$.
Indeed, one has to
update the assertion
$\boundedsrc{\graph}{\spaninitial}$
into
$\boundedsrc{(\{(\vertex_1,\vertex'), (\vertex_2,\vertex')\} \,\cup\, \graph)}{(\blockupd\spaninitial{\vertex'}{c'})}$
for some initial amount of span credits $c'$ for $\vertex'$.
Because of the $\boundedsrcname$ assertion,
this update requires providing at least $c'$ span credits
of $\vertex_1$ and $\vertex_2$, since the new computation
graph is updated with two new edges from $\vertex_1$ and $\vertex_2$ to $\vertex'$~(see \RULE{SchedJoin} in \Cref{fig:semantics}).
Yet, the user is not required to give any span credit
at the completion of a parallel composition,
and thus no credits are available.
Hence, we choose $c' = 0$,
that is, to initially give $0$ span credit to $\vertex'$.
How are the span credits of $\vertex'$ created then?
The answer lies in the proof of the \RULE{PrimitiveTransfer} rule, which we cover next.

\paragraph{Validity of the transfer of span credits}
Recall that \RULE{PrimitiveTransfer} allows for transferring span credits
from a task $\vertex$ to another task $\vertex'$,
provided that the current goal is a WP with current task $\vertex'$,
and that the user has a witness $\cantransfer{\vertex}{\vertex'}$.
From this witness, we deduce that two facts hold:
(1) there is a path from $\vertex$ to $\vertex'$ in the underlying computation graph $\graph$,
and (2) every predecessor of $\vertex'$ in $\graph$
can be reached by $\vertex$.
Moreover, since \RULE{PrimitiveTransfer} must be applied facing
a WP with current task $\vertex'$,
we know that $\vertex'$ has no successor in $\graph$.
From these three facts, we use $n$ span credits of $\vertex$
to generate $n$ more span credits for all
the tasks $\vertex$ can reach (including $\vertex'$),
in cascade.
For every task that has successors,
we store in the \boundedsrcname assertion additional span
credits, in order to justify the validity of
the increase of the span credits of their successors.
Since~$\vertex'$ has no successor, we don't need
to store away the generated $n$ credits in \boundedsrcname,
and we can give them back to the user.

\section{Case Studies}
\label{sec:case_studies}
In this section,
we illustrate the expressiveness of \logic by verifying multiple case studies.
We first present our implementation and specification
of two fundamental building blocks of parallel computations: the parallel for loop~(\cref{sec:parfor})
and the scan operation~(\cref{sec:scan}).
We then devote our attention to parallel sorting,
and verify a parallel merge function~(\cref{sec:parmerge}),
on which we build a parallel merge sort~(\cref{sec:mergesort}).
We finally verify Treiber's stack, a linearizable lock-free stack,
for which we give work and span bounds~(\cref{sec:treiber}).
While we do not prove any tightness results for these bounds,
the costs of all examples are asymptotically as tight
as one can expect from the literature.

Since $\etick$ operations model costs, their placement is critical.
In our case studies,
we place a $\etick$ operation in each branch that performs a
recursive call.

\subsection{The Parallel For Loop}
\label{sec:parfor}
The parallel for loop~$\parforname$ is a fundamental building block of parallel computations.
It allows for executing a given function in parallel on every index of a given range.

\paragraph{Code}
\begin{figure}
\centering\small\morespacingaroundstar
\begin{minipage}{0.4\textwidth}
\begin{align*}
&\parforname\;\eqdef\;\hat\mu\selfname.\,\lambda \lowbound\,\highbound\,\parforarg.\\
&\quad\kw{if}\, \diffname\,\leq\,\zero\,\kw{then}\,\vunit\\
&\quad\kw{else\,if}\,\eeq\diffname\one\,\kw{then}\,\parforarg\,\lowbound\\
&\quad\kw{else}\,\kw{let}\,\midname\,=\, \lowbound + \diffname/2\,\kw{in}\\
&\qquad\etick\sequence\,\epar{(\selfname\,\lowbound\,\midname\,\parforarg)}{(\selfname\,\midname\,\highbound\,\parforarg)}
\end{align*}
\end{minipage}
\hfill
\begin{minipage}{0.59\textwidth}
\begin{mathpar}
\inferrule[Parfor]
{\workc{\highbound - \lowbound - 1} \\
 \spanc{\vertex}{\roundup{\log_2(\highbound - \lowbound)} + C} \\
 \bigsepn{\ofs}{\lowbound}{\highbound}{\forall \vertex'.\;\spanc{\vertex'}{C} \wand \wpgex{\vertex'}{(\parforarg\;\ofs)}{\_}{\_}{\mpost\,\ofs}}}
{\wpgex{\vertex}{(\parfor{\lowbound}{\highbound}{\parforarg})}{\_}{\_}{\bigsepn{\ofs}{\lowbound}{\highbound}{\mpost\,\ofs}}}
\end{mathpar}
\end{minipage}
\captionlabel{The parallel for loop and its specification}{fig:parfor}
\end{figure}

The upper part of \Cref{fig:parfor} presents the code of $\parforname$.
The $\parforname$ function takes three arguments:
a lower bound~$\lowbound$, an upper bound~$\highbound$,
and a function $\parforarg$ to execute in parallel
at every index in $[\lowbound, \highbound)$.
The function is defined recursively, with $\selfname$ as the
recursive name.
It first checks whether $\diffname$ is negative or zero, in which case there is nothing to do, and
then whether $\diffname$ is equal to $1$, in which case the function $\parforarg$ is called on $\lowbound$.
If $\diffname \geq 2$, then the function computes the midpoint $\midname$,
performs a $\etick$ operation, and recursively calls itself on the ranges $[\lowbound, \midname)$ and $[\midname, \highbound)$.
We place a tick operation before the parallel primitive: the cost incurred
by this tick operation accounts for the generation of a binary tree of tasks.

\paragraph{Specification}
The lower part of \Cref{fig:parfor}
presents the specification of a call to
$\parfor{\lowbound}{\highbound}{\parforarg}$ on task $\vertex$.
The work and span costs of $\parforname$
reflect the fact that the execution graph of this function contains a
parallel binary tree with $\diffname$ leaves calling $\parforarg$,
with a $\etick$ operation at every internal node.
As a result,
the first part of the precondition consumes $(\highbound - \lowbound - 1)$
work credits, which upper bounds the number of internal nodes in this tree.
Next, the second premise of the precondition consumes $\roundup{\log_2(\highbound - \lowbound)} + C$
span credits of $\vertex$, where $C$ is a constant selected when applying the rule.
The logarithmic factor accounts for the height of the binary execution tree generated by $\parforname$.
Finally, the third premise of the precondition requires the
user to prove a weakest precondition about every call to $\parforarg$
for an argument~$i$ in $[\lowbound;\highbound)$,
at a universally quantified identifier $\vertex'$,
assuming $C$ span credits are available.
These $C$ credits come from what is put in as part of the second premise, and represent the remaining span credits that reach each leaf before its call to $\parforarg$.
The postcondition returns the iterated conjunction of the postconditions
of the calls to $\parforarg$.

\subsection{Scan (Parallel Prefix Sums)}
\label{sec:scan}
\begin{figure}
\centering\small\morespacingaroundstar
\begin{minipage}{0.34\textwidth}
\begin{align*}
&\tabulatename\;\eqdef\;\lambda \parforarg\,\allocsize.\\
&\quad\kw{let}\,x\,=\,\ealloc{\allocsize}{\vunit}\,\kw{in}\\
&\quad\kw{let}\,k\,=\,\lambda \ofs.\,\estore{x}{\ofs}{(\parforarg\,\ofs)}\,\kw{in}\\
&\quad\parfor{\zero}{\allocsize}{k}\sequence\,x
\end{align*}
\end{minipage}\hfill
\begin{minipage}{0.64\textwidth}
\begin{align*}
&\scanname\;\eqdef\;\hat\mu\selfname.\,\lambda \scanarg.\\
&\quad\kw{let}\,\allocsize\,=\,\elength{\scanarg}\,\kw{in}\\
&\quad\kw{if}\,\eeq{\allocsize}{\one}\,\kw{then}\,\kw{let}\,r\,=\,\ealloc{2}{\zero}\,\kw{in}\,\estore{r}{1}{\eload{\scanarg}{0}}\sequence\,r\,\kw{else}\\
&\quad\kw{let}\,c\,=\,\tabulate{(\lambda \ofs.\,\eload{\scanarg}{2\times\ofs} + \eload{\scanarg}{2\times\ofs+1})}{(\allocsize/2)}\,\kw{in}\\
&\quad\kw{let}\,p\,=\,\etick; \selfname\;c\,\kw{in}\\
&\quad\kw{let}\,g\,=\,\lambda \ofs.\,\kw{let}\,j\,=\,\ofs/2\,\kw{in}\\
&\quad\quad\kw{if}\,\eeq{(\ofs \mathbin{\kw{mod}} 2)}{\zero}\,\kw{then}\,\eload{p}{j}\,\kw{else}\,\eload{p}{j} + \eload{\scanarg}{\ofs-1}\,\kw{in}\\
&\quad\tabulate{g}{(\allocsize+1)}
\end{align*}
\end{minipage}
\begin{mathpar}
\inferrule[Tabulate]
{ \allocsize \neq 0 \\ \workc{\allocsize - 1} \\
 \spanc{\vertex}{\roundup{\log_2 \allocsize} + C} \\
 \bigsepn{\ofs}{0}{\allocsize}{\forall \vertex'.\;\spanc{\vertex'}{C} \wand \wpgex{\vertex'}{(\parforarg\;\ofs)}{\_}{v}{\mpost\,\ofs\,v}}}
{\wpgex{\vertex}{(\tabulate{\parforarg}{\allocsize})}{\_}{v}{\exists \loc\;\wals.\;\pure{v = \loc \;\wedge\; \length{\wals} = \allocsize} \star \loc \mapsto \wals \star \bigsepn{\ofs}{0}{\allocsize}{\mpost\,\ofs\,\wals(\ofs)}}}

\inferrule[Scan]
{\pure{\length{\vals} = 2^k} \\
 \workc{3 \times 2^k} \\
 \spanc{\vertex}{(k+1)^2} \\
 \scanarg \mapsto \vals}
{\wpgex{\vertex}{(\scanname\;\scanarg)}{\_}{v}{\exists \loc\;\wals.\;\pure{v = \loc \;\wedge\; \scanned{\vals}{\wals}} \star \scanarg \mapsto \vals \star \loc \mapsto \wals}}
\end{mathpar}
\vspace{0.1em}
\[\scanned{\vals}{\wals} \;\eqdef\; \length{\wals} = \length{\vals} + 1 \;\wedge\; \forall i.\;0 \leq i \leq \length{\vals} \implies \wals(i) = \sum_{j < i}\,\vals(j)\]
\vspace{-0.5em}
\captionlabel{The tabulate primitive, the scan function, and their specifications}{fig:scan}
\end{figure}

The scan operation, or parallel prefix sum,
is another standard building block of parallel algorithms~\citep{blelloch-prefix-sums}.
Given an array $\vals$ and an associative sum operation,
the scan operation returns another array $\wals$
such that $\wals[0] = 0$ and $\wals[i]$ stores the sum of $\vals[0], \dots, \vals[i-1]$.
Formally,
the last line of \Cref{fig:scan} defines
an assertion $\scanned{\vals}{\wals}$ stating that $\wals$ stores the result of the scan of array $\vals$.

While the scan operation might at first seem to be inherently sequential, it
can be efficiently implemented in parallel.
In this section, for simplicity, we focus on arrays of integers
that have a size that is a power of 2.

\paragraph{Code}
The upper part of \Cref{fig:scan}
first presents the code of
the $\tabulatename$ function,
a utility function that takes as argument
a function $\funcname$ and an integer $\allocsize$,
and which constructs a fresh array
of size~$\allocsize$ whose $\ofs$-th entry is the result of
$\funcname\;\ofs$, computed in parallel using $\parforname$.
The upper part of \Cref{fig:scan}
then presents the $\scanname$ function itself,
which implements a contraction algorithm,
similar to the standard three-phase algorithm~\cite{DBLP:conf/sc/ChatterjeeBZ90}.
This function takes as argument
an array $\scanarg$ and is defined recursively,
with recursive name $\selfname$.
The function first
tests if the argument contains a single element,
in which case it returns a new array of size 2,
where the first index stores~0 and the second index stores
the single element of $\scanarg$.
Otherwise, the function has three phases.
First, it allocates the contraction $c$,
which is an array of half the size of the argument,
which stores the sum of each pair of consecutive elements.
Second, it performs a tick operation and calls itself recursively to compute the
prefix sum of this intermediate array~$c$ and names
the result $p$.
Third, it expands the result back to full size
using a second tabulate:
even-indexed entries are taken directly from $p$,
while odd-indexed entries are each obtained by adding an input element
to the preceding even-index prefix sum.
As we will see, our implementation of $\scanname$
has $\complex{n}$ work and $\complex{\log^2(n)}$ span, where $n$ is the length of the input array.
Our implementation of $\scanname$ has a tick operation before the recursive call,
accounting for the (logarithmic) depth of the recursion.
Note that other costs will be incurred by
calls to $\tabulatename$.

\paragraph{Specification}
The lower part of \Cref{fig:scan} first presents the specification of
a call to $\tabulate{\parforarg}{\allocsize}$ on task~$\vertex$.
Its specification is similar to the one of $\parforname$.
It consumes $\allocsize - 1$ work credits
and $\roundup{\log_2 \allocsize} + C$ span credits of~$\vertex$,
and requires the user to verify every call to~$\parforarg$
for an argument~$\ofs$ in $[0;\allocsize)$,
at a universally quantified identifier~$\vertex'$,
assuming $C$ span credits are available.
The postcondition asserts that the function returns
an array~$\wals$ of length~$\allocsize$,
such that $\mpost\;\ofs\;\wals(\ofs)$ holds
for every $\ofs \in [0;\allocsize)$.

The lower part of \Cref{fig:scan} then presents
the specification of
a call to $\scanname\,\scanarg$ on task~$\vertex$.
The precondition
requires that $\scanarg$ points to
an array $\vals$ of size $2^k$.
It also consumes $3 \times 2^k$ work credits
and $(k+1)^2$ span credits of~$\vertex$,
yielding linear work and polylog span.
The postcondition asserts that the function returns
a fresh array $\loc$ with the prefix sums of the argument $\scanarg$,
which is returned untouched.
Instead of working directly with closed-form
expressions for the work and span,
we establish the proof using open forms in terms of recurrence relations.
More precisely, we first define
(at the logical level) recursive functions $\scanwk(n)$ and $\scansp(n)$, which
will represent
the total work and span credits, respectively, that will be needed for an input of length $n$
(since $n$ is always a power of two, we ignore rounding in division):
\begin{align*}
\scanwk(n) &\eqdef \text{if } n \leq 1 \text{ then } 0 \text{ else } (n/2 - 1) + 1 + \scanwk(n/2) + n \\
\scansp(n) &\eqdef \text{if } n \leq 1 \text{ then } 0 \text{ else } \roundup{\log_2(n/2)} + 1 + \scansp(n/2) + \roundup{\log_2 (n+1)}
\end{align*}
For both work and span, the non-base case is a sum of four terms.
The first term accounts for the cost of the first tabulate,
which allocates an array of size $n/2$ and for which the executed closure has no cost (the constant $C$ in \RULE{Tabulate} is instantiated to $0$).
The second term accounts for the tick operation before the recursive call.
The third term accounts for the recursive call.
The fourth term accounts for the cost of the second
tabulate, which allocates an array of size $n+1$ with again a free closure.
Working in terms of these recurrence relations,
we split up the parts of the non-base case to pay for each of these respective components.
Outside of the logic, we then separately establish closed upper bounds on the recurrence relation showing that
$\scanwk(n) \leq 3 \times n$
and $\scansp(n) \leq (\roundup{\log_2(n)} + 1)^2$.
Since \logic{} is affine, we can use \RULE{WorkSplit} and \RULE{SpanSplit} to weaken our specification to the form shown in \RULE{Scan} with these upper bounds.

The proof makes crucial use of \RULE{Transfer}.
Indeed,
before executing the first tabulate operation,
we split span credits in two, one assertion with
$\log_2(n/2)$ for the upcoming tabulate,
and $(\scansp(n/2) + \log_2 (n+1))$ for the remaining computation.
However, because tabulate will change the task identifier,
we have to use \RULE{Transfer} in order to transfer the second
assertion to the continuation.

\subsection{Parallel Merge}
\label{sec:parmerge}
The parallel merge sort that we will present
in \Cref{sec:mergesort}
relies on a fundamental component:
the ability to merge two sorted arrays in parallel.
In this section, we first show how to verify such a parallel merge.
The code we verify is a direct
translation into \lang of the implementation
of merge from the MaPLe standard library~\citep{mpllib}.

\paragraph{Slices}
In order to preserve a relatively
concise definition for $\mergename$,
the function doesn't work directly on arrays but on \emph{slices}.
A slice is a tuple
of an array location, a lower index and an upper index,
describing a portion of the underlying array.
Above this wrapper, we define functions \slengthname,
\sloadname, \sstorename as expected.
We also define the function \ssplitname,
which extracts a sub-slice between two indices.
Note that \ssplitname does not allocate a new array---the
returned slice shares the same underlying array as its argument.
All of these operations are constant time and have no tick operations.
At the specification level,
slices are represented with the $\sliceown{\loc}{\vals}$
predicate, which behaves close to the standard points-to predicate.
This predicate is implemented with standard ghost state,
which we omit for brevity.

\paragraph{Code}
\begin{figure}
\centering\small\morespacingaroundstar
\begin{align*}
&\mergename\;\eqdef\;\hat\mu\selfname.\,\lambda g\;s_1\;s_2\;d.\\
&\quad\kw{if}\,\slength{d}\le g\,\kw{then}\,\smergeseq{s_1}{s_2}{d}\\
&\quad\kw{else}\,\kw{let}\,n_1\,=\,\slength{s_1}\,\kw{in}\,\kw{let}\,n_2\,=\,\slength{s_2}\,\kw{in}\\
&\qquad\kw{if}\,\eeq{n_1}{\zero}\,\kw{then}\,\scopyseq{s_2}{d}\\
&\qquad\kw{else}\,\kw{let}\,\midname_1\,=\,n_1/2\,\kw{in}\\
&\qquad\quad\kw{let}\,p\,=\,\sload{s_1}{\midname_1}\,\kw{in}\\
&\qquad\quad\kw{let}\,\midname_2\,=\,\binsearch{s_2}{p}\,\kw{in}\\
&\qquad\quad\kw{let}\,l_1\,=\,\ssplit{s_1}{0}{\midname_1}\,\kw{in}\,\kw{let}\,r_1\,=\,\ssplit{s_1}{(\midname_1+1)}{n_1}\,\kw{in}\\
&\qquad\quad\kw{let}\,l_2\,=\,\ssplit{s_2}{0}{\midname_2}\,\kw{in}\,\kw{let}\,r_2\,=\,\ssplit{s_2}{\midname_2}{n_2}\,\kw{in}\\
&\qquad\quad\sstore{d}{(\midname_1+\midname_2)}{p};\\
&\qquad\quad\kw{let}\,d_l\,=\,\ssplit{d}{0}{(\midname_1+\midname_2)}\,\kw{in}\,\kw{let}\,d_r\,=\,\ssplit{d}{(\midname_1+\midname_2+1)}{(\slength{d})}\,\kw{in}\\
&\qquad\quad\etick;\,\epar{(\selfname\;g\;l_1\;l_2\;d_l)}{(\selfname\;g\;r_1\;r_2\;d_r)}
\end{align*}
\begin{mathpar}
\inferrule{}
{\mergewk(n, m) \;\eqdef\; 11 \times (n+m)}

\inferrule{}
{\mergesp(g, n, m) \;\eqdef\; g + 4 \times (\logtwoup{n} + 1) \times (\logtwoup{m} + 1)}

\inferrule[Merge]
{\pure{\sorted{\vals_1} \;\wedge\; \sorted{\vals_2} \;\wedge\; \length{\wals} = \length{\vals_1} + \length{\vals_2}} \\
 \workc{\mergewk(\length{\vals_1}, \length{\vals_2})} \\
 \spanc{\vertex}{\mergesp(g, \length{\vals_1}, \length{\vals_2})} \\
 \sliceown{s_1}{\vals_1} \\ \sliceown{s_2}{\vals_2} \\ \sliceown{d}{\wals}}
{\wpgex{\vertex}{(\mergename\;g\;s_1\;s_2\;d)}{\_}{v}{\sliceown{s_1}{\vals_1} \star \sliceown{s_2}{\vals_2} \star \sliceown{d}{(\mergeop{\vals_1}{\vals_2})}}}
\end{mathpar}
\captionlabel{The parallel merge operation and its specification}{fig:parmerge}
\end{figure}

The upper part of \Cref{fig:parmerge} presents the code
of \mergename.
The function takes four arguments:
a granularity threshold~$g$,
two sorted source slices~$s_1$ and~$s_2$,
and a target slice~$d$.
The function is defined recursively,
with $\selfname$ as the recursive name,
and merges $s_1$ and $s_2$ into $d$, overwriting its content.
If $\slength{d} \leq g$, the function falls back to a sequential merge (code omitted).
If $s_1$ is empty,
the function copies~$s_2$ into~$d$.
Otherwise, it selects the median element of~$s_1$ as a pivot~$p$,
and uses binary search to find the index~$\midname_2$
in~$s_2$ such that all elements before~$\midname_2$ are
smaller than~$p$ and all elements after are greater.
It places~$p$ at its final position~$\midname_1 + \midname_2$ in~$d$,
and splits all three slices around this pivot position.
After a $\etick$ operation, the function
recursively merges the left and right halves in parallel.
By placing a tick operation before the parallel primitive,
we account for the creation of a binary tree of tasks.

\paragraph{Specification}
Before diving into the work and span bounds of $\mergename$,
we first present bounds for intermediate sequential functions---for which work and span are equal---%
assuming $n$ is the length of the first array $s_1$
and $m$ is the length of the second array $s_2$.
\[\addtolength{\arraycolsep}{0.3em}\begin{array}{c|c|c|c}
\text{Function} & \smergeseq{s_1}{s_2}{d} & \scopyseq{s_2}{d} & \binsearch{s_2}{p} \\\hline
\text{Work and Span} & n+m & m & \roundup{\log_2(m)} + 1
\end{array}\]

The center part of \Cref{fig:parmerge}
presents the work and span bounds we derive for \mergename.
The function $\mergewk(n, m)$ gives the work of $\mergename$
when the first argument has length $n$ and second argument has length $m$,
and states that the work is in $\complex{n+m}$.
The function $\mergesp(g,n,m)$ presents the span, which is in
$\complex{g + \log_2(n)\times\log_2(m)}$.
Indeed, \mergename calls a binary search
on the second array, which has logarithmic span,
at every recursive call, and the depth of the recursion is logarithmic.
As for \scanname~(\cref{sec:scan}),
we obtained these bounds by first conducting the proof
using open recurrence equations
\bgroup\normalsize\begin{align*}
\mergewkeq(g,n, m) \eqdef\;&
\text{if } n + m \leq g \text{ then } n + m \text{ else if } n = 0 \text{ then } m - 1 \\
&\text{else } 1 + (1 + \roundup{\log_2 m}) + \\
&\phantom{\text{else }}\max_{0 \leq p \leq m}\big(\mergewkeq(g,\lfloor n/2 \rfloor, p) + \mergewkeq(g,n {-} 1 {-} \lfloor n/2 \rfloor, m {-} p)\big)\\
\mergespeq(g,n, m) \eqdef\;&
\text{if } n + m \leq g \text{ then } n + m \text{ else if } n = 0 \text{ then } \roundup{\log_2 m} \\
&\text{else } (1 + \roundup{\log_2 m}) + 1 + {} \\
&\phantom{\text{else }}\max_{0 \leq p \leq m}\big(\max\;(\mergespeq(g,\lfloor n/2 \rfloor, p))\;(\mergespeq(g,n {-} 1 {-} \lfloor n/2 \rfloor, m {-} p))\big)
\end{align*}\egroup
We then show in Rocq that $\mergewkeq(g,n, m) \leq \mergewk(n, m)$ and $\mergespeq(g,n, m) \leq \mergesp(g, n, m)$.
Formally proving a closed form for the work was a difficult task; we posit
that we are the first to give a machine-checked proof of this.

The lower part of \Cref{fig:parmerge}
presents the formal specification of a call to $\mergename\,g\,s_1\,s_2\,d$
on task~$\vertex$.
The precondition requires the two source slices
$s_1$ and $s_2$ to represent arrays $\vals_1$ and $\vals_2$,
which must be sorted,
and the target slice $d$ to represent some irrelevant array $\wals$
whose length equals $\length{\vals_1} + \length{\vals_2}$.
It also consumes work credits $\mergewk(\length{\vals_1}, \length{\vals_2})$
and span credits $\mergesp(g, \length{\vals_1}, \length{\vals_2})$.
The postcondition returns ownership of the source slices unchanged,
and asserts that the destination slice~$d$ now contains $\mergeop{\vals_1}{\vals_2}$,
the sorted merge of $\vals_1$ and $\vals_2$.

\subsection{Parallel Merge Sort}
\label{sec:mergesort}
\newcommand{\sameelems}[2]{\textsf{permutation}\,#1\,#2}
\begin{figure}
\centering\small\morespacingaroundstar
\begin{align*}
&\sortname\;g\;\eqdef\;\hat\mu\selfname.\,\lambda b\;s\;d.\\
&\quad\kw{let}\,n\,=\,\slength{s}\,\kw{in}\\
&\quad\kw{if}\,\eeq{n}{\zero}\,\kw{then}\,\vunit\\
&\quad\kw{else\,if}\,\eeq{n}{\one}\,\kw{then}\,\kw{if}\,b\,\kw{then}\,(\sstore{d}{\zero}{(\sload{s}{\zero})})\,\kw{else}\,\vunit\\
&\quad\kw{else}\,\kw{let}\,\midname\,=\,n/2\,\kw{in}\\
&\qquad\kw{let}\,s_l\,=\,(\ssplit{s}{\zero}{\midname})\,\kw{in}\,\kw{let}\,s_r\,=\,(\ssplit{s}{\midname}{n})\,\kw{in}\\
&\qquad\kw{let}\,d_l\,=\,(\ssplit{d}{\zero}{\midname})\,\kw{in}\,\kw{let}\,d_r\,=\,(\ssplit{d}{\midname}{n})\,\kw{in}\\
&\qquad\etick\sequence\,\epar{(\selfname\,\neg b\,s_l\,d_l)}{(\selfname\,\neg b\,s_r\,d_r)}\sequence\\
&\qquad\kw{if}\,b\,\kw{then}\,(\mergename\,g\,s_l\,s_r\,d)\,\kw{else}\,(\mergename\,g\,d_l\,d_r\,s)
\end{align*}
\begin{mathpar}
\inferrule{}
{\sortwk(n) \;\eqdef\; 12 \times n \times \logtwoup{n}}

\inferrule{}
{\sortsp(g, n) \;\eqdef\; g + (g + 5) \times (\logtwoup{n} + 1)^3}

\inferrule[Sort]
{\pure{\length{\vals} = \length{\wals}} \\
 \workc{\sortwk(\length{\vals})} \\
 \spanc{\vertex}{\sortsp(g, \length{\vals})} \\
 \sliceown{s}{\vals} \\ \sliceown{d}{\wals}}
{\wpgexlong{\vertex}{(\sortname\;g\;b\;s\;d)}{\_}{v}{%
  \begin{array}{@{}l@{}}
    \exists \vals'\,\wals'.\;
    \pure{\sorted{\vals'} \;\wedge\; \sameelems{\vals'}{\vals} \;\wedge\;
      \length{\wals'} = \length{\wals}} \star {} \\
    \sliceown{s}{(\kw{if}\;b\;\kw{then}\;\wals'\;\kw{else}\;\vals')} \star
    \sliceown{d}{(\kw{if}\;b\;\kw{then}\;\vals'\;\kw{else}\;\wals')}
  \end{array}}}
\end{mathpar}
\captionlabel{The parallel merge sort function and its specification}{fig:merge_sort}
\end{figure}

The parallel merge sort we verify is
a direct translation into \lang of the implementation
of merge sort from the MaPLe standard library~\citep{mpllib}.

\paragraph{Code}
The code appears in the upper part of \Cref{fig:merge_sort},
and uses a ping-pong/double buffering approach
to remove the need for copies.
The function is parameterized by
a granularity threshold~$g$,
a boolean $\vbool$,
a source array $s$, and a destination array $d$ of the same size.
When $\vbool = \vtrue$, the function writes the sorted contents of $s$ into $d$,
leaving~$s$ with garbage content.
When $\vbool = \vfalse$, the function sorts $s$ in-place,
using $d$ as temporary memory.
The function \sortname is defined recursively,
with $\selfname$ as the recursive name.
If $s$ is empty, there is nothing to do.
If $s$ has a single element, it is already sorted;
the function copies this single element to $d$ if
$\vbool = \vtrue$, and does nothing otherwise.
In the general case, the function computes the midpoint $\midname$,
splits both $s$ and~$d$ into left and right halves around~$\midname$,
and after a $\etick$ operation,
recursively sorts the two halves in parallel.
The boolean argument is flipped for the recursive calls, alternating the buffer.
Finally, the function merges the two sorted halves:
if $\vbool = \vtrue$, the halves of~$s$ are merged into~$d$;
and otherwise, if $\vbool = \vfalse$,
the halves of~$d$ are merged into~$s$.
Similarly to $\parforname$ and $\mergename$,
by placing a tick operation before the parallel primitive,
we account for the creation of a binary tree of tasks.

\paragraph{Specification}
The work and span bounds of \sortname are given in the center part of \Cref{fig:merge_sort}.
We show that \sortname has work in $\complex{n \log n}$
and that its span is in $\complex{\log^3 n}$, where $n$ is the length of the input array.
These bounds are again obtained by closing open recurrence equations---we refer
the curious reader to our formalization~\citep{mechanization}.

The specification of
a call to $\sortname\,g\,b\,s\,d$ on task $\vertex$
appears in the lower part of \Cref{fig:merge_sort}.
The precondition requires that $s$ is a slice of content $\vals$
and $d$ a slice of content $\wals$, of the same length.
The precondition also consumes the expected work and span credits.
The postcondition asserts that there
exist two new arrays.
First $\vals'$, the sorted content of $\vals$---this is witnessed by
the property $\sorted{\vals'}$ that says that $\vals'$ itself is sorted,
and $\sameelems{\vals'}{\vals}$,
that says that $\vals'$ contains a permutation of $\vals$.
Second $\wals'$ is the content of the buffer that was used,
that is, an array with unspecified content but of the same length
as the arguments.
The content of the two slices $s$ and $d$ depends on
the value of the boolean $b$.
If $b = \vtrue$, the sorted array $\vals'$ is stored in $d$
and $s$ is left with unspecified content~$\wals'$;
if $b = \vfalse$, the sorted array $\vals'$ is stored in $s$
and $d$ is left with unspecified content~$\wals'$.

\subsection{Treiber's Stack}
\label{sec:treiber}
\newcommand{\atriple}[6]{%
\begin{array}{c}
\{#3\}\arcr
\langle #4 \rangle\arcr
#1 : #2 \arcr
\langle #5 \rangle\arcr
\{#6\}
\end{array}%
}

\newcommand{\atripleinline}[6]{%
\{#3\}\langle#4\rangle\,#1 : #2\,\langle#5\rangle\{#6\}
}

\begin{figure}\centering\small\morespacingaroundstar
\begin{minipage}{0.48\textwidth}
\begin{align*}
&\stackcreatename\;\eqdef\;\lambda \_.\,\ealloc{1}{\vunit}\\[6pt]
&\stackpushname\;\eqdef\;\hat\mu\selfname.\,\lambda s\,v.\\
&\quad\kw{let}\,h'\,=\,\ealloc{2}{v}\,\kw{in}\\
&\quad\kw{let}\,h\,=\,\eload{s}{0}\,\kw{in}\\
&\quad\estore{h'}{1}{h}\sequence\\
&\quad\kw{if}\,\ecas{s}{0}{h}{h'}\,\kw{then}\,\vunit\,\kw{else}\,(\etick\sequence\,\selfname\,s\,v)
\end{align*}
\end{minipage}\hfill
\begin{minipage}{0.48\textwidth}
\begin{align*}
&\stackpopname\;\eqdef\;\hat\mu\selfname.\,\lambda s.\\
&\quad\kw{let}\,h\,=\,\eload{s}{0}\,\kw{in}\\
&\quad\kw{if}\,\eeq{h}{\vunit}\,\kw{then}\,\vunit\,\kw{else}\\
&\qquad\kw{let}\,h'\,=\,\eload{h}{1}\,\kw{in}\\
&\qquad\kw{if}\,\ecas{s}{0}{h}{h'}\,\kw{then}\,\eload{h}{0}\,\kw{else}\,(\etick\sequence\,\selfname\,s)
\end{align*}
\end{minipage}
\begin{mathpar}
\inferrule[Create]
{ 0 \leq n \\ 0 \leq m}
{\wpgex{\vertex}{(\stackcreatename\;\vunit)}{\_}
{v}
{\exists s.\;\pure{v = s} \star \stackown{s}{n}{m}{[]} \star \bigsepnbase{}{0}{n}{(\forall\vertex'.\; \spanc{\vertex'}{m} \wand \player{s}{\vertex'})}}}

\inferrule[ParticipantTransfer]
{\player{s}\vertex \\
  \wpgex{\vertex}{\expr}{\vertex'}{\val}{ \player{s}{\vertex'} \wand \post\,\vertex'\,\val}}
{\wpg{\vertex}{\expr}{\post}}
\\
\inferrule[StackPush]{}
{\atriple{\vertex}{(\stackpush{s}{x})}
{\player{s}\vertex}
{\forall m\,xs.\,\workc{n-1} \star \stackown{s}{n}{(m+1)}{xs}}
{\stackown{s}{n}{m}{(x::xs)}}
{\lambda\vertex'\,\_.\;\pure{\vertex'=\vertex} \star \player{s}{\vertex}}}

\inferrule[StackPop]{}
{\atriple{\vertex}{(\stackpop{s})}
{\player{s}\vertex}
{\forall m\,x\,xs.\,\workc{n-1} \star \stackown{s}{n}{(m+1)}{(x::xs)}}
{\stackown{s}{n}{m}{xs}}
{\lambda\vertex'\,\val.\;\pure{\vertex'=\vertex \wedge \val=x} \star \player{s}{\vertex}}}
\end{mathpar}
\captionlabel{Code and specifications for Treiber's stack}{fig:treiber}
\end{figure}

Treiber's stack~\citep{treiber-86}
is a concurrent, linearizable~\citep{herlihy-wing-90}, and lock-free stack.
With \logic, we show that indeed Treiber's stack is linearizable
(in a style similar to those in prior Iris-based logics),
and that it satisfies a performance property:
assuming a fixed number of participants,
we provide work bounds for push and pop,
meaning that multiple push and pop operations can happen in parallel,
and none of them can be indefinitely delayed by the others.
Moreover, the span bounds we provide for push and pop intuitively
illustrate the worst case cost of contention.

\paragraph{Code}
The upper part of \Cref{fig:treiber} presents
our implementation of Treiber's stack.
The stack is represented as a reference on a
linked list of nodes.
Each node is an array of size 2, where the first index stores the value of the node and the second index stores a pointer to the next node.
The nil case is represented by the unit value~$\vunit$.
The function $\stackcreatename$
makes a new stack by allocating a reference to an empty list.
The function call $\stackpush{s}{x}$ pushes a value~$x$ on the stack~$s$.
This function first
allocates a new node,
writes $x$ and the current head of the stack in this node,
and then performs an atomic CAS operation
to try to update the reference $s$ to this new cell.
If the CAS succeeds, the function returns,
and otherwise it calls itself recursively to make another attempt.
The function call $\stackpop{s}$
pops the head of the stack $s$.
The function first loads the current content of $s$ into $h$.
If $h = \vunit$, the stack is empty and the call returns.
Otherwise, we know that $h$ is a proper list node,
the function then tries to update $s$ to
the next node using an atomic CAS operation.
If the CAS succeeds,
the function returns the value of the head of the stack stored in $h$,
and otherwise it calls itself recursively to make another attempt.
For both \stackpushname and \stackpopname,
the tick operation accounts for the cost of a failed CAS operation,
that is, for the cost of contention.

\paragraph{Atomic triples}
Our specifications of push and pop
specify that these functions are linearizable.
To express this property, we use an \emph{atomic triple}~\citep{da-rocha-pinto-et-al-14,iris-15},
which indeed guarantees linearizability~\citep{birkedal-et-al-21}.
In our work,
an atomic triple takes
the form
\[\atripleinline{\vertex}{\expr}{\pre_1}{\forall x.\,\pre_2}{\pre_3}{\lambda \vertex'\,\val.\;\pre_4}\]
Such a triple specifies the execution
of $\expr$ on task $\vertex$.
The precondition $\pre_1$ is called the \emph{private precondition}
and must hold at the beginning of the execution of $\expr$.
The precondition $\pre_2$ is called the \emph{public precondition}
and is guaranteed to be updated atomically,
at some point during the execution of $\expr$,
to the public postcondition $\pre_3$.
Finally, the private postcondition $\pre_4$
holds at the end of the execution of $\expr$.
Note the universal quantifier
in the public precondition, which scopes over the public precondition,
and both the public and private postconditions.
Our encoding of atomic triples follows the standard Iris recipe~\citep{irislogatom}.

\paragraph{Specification}
The lower part of \Cref{fig:treiber} presents the specification of the three functions \stackcreatename, \stackpushname, and \stackpopname.
These specifications involve two new assertions.
First, the assertion $\stackown{s}{n}{m}{xs}$ states
that $s$ is a stack that may be used concurrently by at most
$n$ participants to perform at most $m$
operations,
and that the content of the stack is the list of values $xs$.
Second,
the assertion $\player{s}{\vertex}$ states that
$\vertex$ is a participant for stack $s$, that is,
a task that may perform operations on~$s$.
These two assertions are not duplicable.
Moreover, the assertion $\player{s}{\vertex}$
behaves as a storage of span credits.
As such, we offer \RULE{ParticipantTransfer}
that mimics \RULE{Transfer} (and is derived from it),
and allows for transferring
a \playername assertion to a subsequent task.

The figure presents the specification of $\stackcreatename\,\vunit$.
The postcondition asserts that the returned value is
a location $s$ that represents
a stack with at most $n$ participants
and $m$ operations ($n$ and~$m$ being chosen by the user),
and that the content of the stack is empty.
Moreover, the postcondition asserts
an $n$-fold iterated conjunction
of separation implications, each producing
a \playername assertion for a given task $\vertex'$
in exchange for $m$ span credits of $\vertex'$.
Indeed, the intention is to bound the span
of every task~$\vertex'$ that performs an operation on the stack by $m$.
Yet, span credits for $\vertex'$ are not yet available
at the moment of the call to $\stackcreatename$, since $\vertex'$ likely
does not even exist yet.

The figure then presents the
specification of $\stackpush{s}{x}$
on task $\vertex$.
The private precondition
requires $\vertex$ to be a declared participant of $s$
as it consumes the assertion $\player{s}{\vertex}$.
The public precondition
universally quantifies
over the remaining number of operations $m$
and the content of the stack $xs$,
and consumes the assertion $\stackown{s}{n}{(m+1)}{xs}$
as well as $(n-1)$ work credits.
The $m+1$ ensures that at least one operation
is still allowed.
The $(n-1)$ work credits are intuitively
used to ``compensate'' for
the success of the CAS operation.
Indeed, there are at most $n-1$ other participants
(executing either push or pop) concurrently.
These participants may have already read the content of $s$
and may themselves be planning to perform a CAS operation to update $s$.
However, because the CAS operation of $\vertex$ succeeded,
the other participants are bound to fail their CAS,
and as such will have to pay for a $\etick$ operation.
The public postcondition
asserts that $x$ has been inserted in the stack,
and that one less operation is allowed.
The private postcondition gives
back the $\player{s}{\vertex}$ assertion,
allowing for a further call to push or pop by $\vertex$.

The specification of $\stackpop{s}$ is similar.
Indeed, the private precondition
consumes a $\player{s}{\vertex}$ assertion.
The public precondition universally quantifies
over the remaining number of operations $m$,
the top of the stack $x$ and the content $xs$;
it then consumes $\stackown{s}{n}{(m+1)}{(x::xs)}$
as well as $(n-1)$ work credits.
The work credits are used in the same way as for push.
The public postcondition asserts that $x$ has been popped from the stack,
and that one less operation is allowed.
The private postcondition
asserts that the returned value is $x$ and gives back the $\player{s}{\vertex}$ assertion.

\paragraph{Proof}
We present the proof technique
we use for the credits consumption,
and refer the curious reader to our mechanization for more details~\citep{mechanization}.

Span credits are handled directly with the $\playername$ assertion.
\RULE{Create} globally fixes $m_0$, the maximum number of push/pop operations,
and we track the number $d$ of operations already made
(both are constrained using ghost state).
The assertion $\player{s}{\vertex}$ stores $(m_0 - d)$ span credits for~$\vertex$,
which suffice to pay for the $\etick$ of each failed CAS.

Work credits are more subtle.
Following the literature on lock-freedom proofs,
a task performing a push/pop operation pays for the work it may cause parallel tasks to incur.
The protocol is enforced by the $\stackownname$ invariant using two ghost tokens.
A ``passive'' token, stored in the $\playername$ assertion, denotes
the capability to perform an operation; an ``active'' token, tagged with a location~$\loc$,
denotes a task currently performing an operation assuming the head of the list is $\loc$.
The invariant stores one work credit per active token whose assumed head is outdated.
A task reading the head exchanges its passive token for an active one tagged with the location read.
Upon a CAS, it accesses the public precondition to obtain $n-1$ work credits,
where $n$ is the number of participants.
If the CAS succeeds, it stores one credit for each of the (at most $n-1$) other tasks
whose active tokens are now outdated, exchanges its active token for a passive one, and returns.
If the CAS fails, it returns the $n-1$ credits to the user, uses its passive token to reclaim
the one credit stored by the contending task that succeeded, pays for the $\etick$, and retries.

\section{Related Work}
\label{sec:related_work}
\paragraph{Program logics for resource usage}
Program logics
for reasoning about resource usage date back to
\citet{nielsonphd, nielsonmore},
who proposes a logic for
the time complexity
of pure programs with while loops.
This logic has no notion of credits.
The idea of credits has its roots
in type systems for time complexity~\citep{hofmann-99, pilkiewicz-pottier-monotonicity-11},
and was introduced in separation logic by~\citet{atkey-11}.
\citet{haslbeck-nipkow-18} compare a specialized Hoare logic
and a separation logic with credits to verify time bounds.

Time credits in separation logic have been used extensively
for verifying the time complexity of
sequential programs.
For instance, \citet{chargueraud-pottier-15, chargueraud-pottier-uf-sltc-19}
use time credits to verify
the time complexity of a union-find implementation,
using the CFML framework~\citep{cfml}.
In the same framework,
\citet{moine-chargueraud-pottier-22}
verify the time complexity of a transient data structure.
\citet{mevel-jourdan-pottier-19}
bring time credits to the Iris world.
They also introduce the dual notion of \emph{time receipts},
for proving a lower bound on the time complexity of a program.
We posit that these receipts could also be adapted for work and span.
Our specifications with work and span credits are
extremely precise~(\cref{sec:case_studies}),
with explicated constants instead of big-O notations.
\citet{gueneau-chargueraud-pottier-18} present
a way of encoding big-O notations in separation logic with time credits,
and we posit that their approach could be adapted to our setting.
More recently, \citet{pottier-et-al-24}
apply time credits in Iris to
programs using thunks to achieve good amortized time complexity.
Standard time credits are well suited for amortization:
one can surcharge the actual cost of an operation,
and store the difference inside another assertion,
prepaying a part of the totality of a future operation.
As our case study on Treiber's stack shows~(\cref{sec:treiber}),
work credits are also well suited for amortization.
Span credits, however, are annotated with task identifiers,
and are thus less amenable to amortization.
This is not a problem in practice,
since span analysis is very rarely amortized.

All the tools presented so far are for reasoning about
\emph{worst case} time complexity.
\citet{tachis} present Tachis, a separation logic
for reasoning about the \emph{expected} cost (and in particular expected time)
of probabilistic programs.
When reasoning on a sampling instruction, Tachis
allows for distributing credits
between the different outcomes, as long as the expected value is preserved.

Credits have also been used for reasoning about other resources.
For instance, \citet{madiot-pottier-22} and \citet{moine-chargueraud-pottier-23}
use space credits to verify the heap space complexity of
programs under garbage collection.
Subsequently, \citet{irisfit} scaled their approach to a concurrent setting.
\citet{eris} and \citet{coneris} propose \emph{error credits},
a logical tool for bounding the error
probability of a probabilistic program.
Credits are also useful as an internal logical tool, as in
\emph{later credits}, which simplify reasoning about the so-called later modality~\citep{spies-et-al-22}.

\paragraph{Type systems for work and span}
Many type systems have been developed for sequential
time complexity, some of which we alluded to in the introduction.
We focus here on the closest related work,
concerning work and span.
\citet{hoffmann-shao-parallel} present a
type system with an inference algorithm to bound
the work and span in a pure, parallel,
and first-order programming language.
In such a restricted setting, they can automatically
infer polynomial bounds (their tool does not support logarithmic factors).
In their type system, types are equipped with a potential,
which can be thought as a bag of credits.
Similarly to us, they note the unsoundness of allowing
unrestricted duplication of potentials at the par rule site.
Contrary to us, they chose a different approach, without duplication.
In their system, at the application of the par rule,
they type check each sub-expression twice,
with potentially different context-splitting of potentials,
but with the same result type (and hence same potential) for each typing.
The first typing is made with typing rules in which operations
incur cost, whereas
the second is made using a ``cost-free'' semantics,
in which potentials are never consumed.
The soundness argument comes from the fact that at runtime,
one of the two sub-expressions will generate the heaviest path,
and hence, one of the pairs of typing derivations
(one with cost, and the other cost-free)
can be used to validate the original bound.
Such an approach is not desirable in a program logic,
where we want to verify a program only once.

\citet{baillot-ghyselen-24} propose a type system
with sized types for the work and span of programs
written in the Pi-calculus, a formal model of parallelism
without shared memory but in which tasks (or processes)
can exchange data using channels.
Their approach is more direct than ours:
in their system,
the span of the parallel composition
of two processes is computed
as the maximum of the span of each process;
whereas in \logic, the duplication of span credits
implicitly encodes the maximum operation.

\citet{calf} and \citet{decalf} present Calf and Decalf,
cost-aware logical frameworks based on type theory,
which allow for proving refinements between
program behaviors as well as between their costs.
Calf supports parallelism and offers
a way to reason about the work and span.
However, because they work outside of separation logic,
they focus on purely functional code and verify pure variants of some of the
parallel algorithms we consider in this paper.
\citet{calf} verify
the work and span of a parallel merge sort implementation,
similar to ours~(\cref{sec:mergesort}).
In unpublished work, \citet{zhou-calf} verifies
several implementations of the scan primitive,
and in particular a contraction-based implementation similar to ours~(\cref{sec:scan}).

\citet{das-work-session} propose
a session-typed system for bounding the work
(in their setting, the total number of messages exchanged).
They scale their approach to the span in a later paper~\citep{das-parallel-session}.
\paragraph{Performance properties of concurrent programs}
In \Cref{sec:treiber},
we give an API to Treiber's stack in \logic.
Our specifications for \stackpushname and \stackpopname
show some performance property.
While this is not a formal proof of lock-freedom,
it is close to it,
and we draw inspiration from prior work.
In particular,
\citet{hoffmann-et-al-13} propose
a separation logic for proving lock-freeness
using a \emph{quantitative compensation scheme},
which ``ensures that a
thread is compensated for loop iterations that are caused by
progress [...] in another thread''.
This is related to the pattern followed by our specifications
for \stackpushname and \stackpopname,
which requires $n-1$ work credits ($n$ being the number of participants),
in order to compensate them for a potential incurred recursion.
\citet{jia-li-vafeiadis-15} propose another approach,
more amenable to automation,
using ghost variables to witness that
if a task failed to make progress, then another task has made progress.

\section{Conclusion and Future Work}
\label{sec:conclusion}
We present \logic,
a separation logic for
proving bounds on the work and span of parallel programs.
We use work credits, akin to standard time credits,
to account for work,
and span credits, a new kind of credits that
are duplicated upon a par operation,
to account for span.
Crucially, span credits are tagged with a
logical task identifier, restricting their use.
We support a key \RULE{Transfer} rule for
transferring span credits between tasks when possible.
We prove the soundness of \logic and illustrate it
with several case studies, including a higher-order
parallel for loop, as well as a parallel merge sort
algorithm and Treiber's lock-free stack.
All results are mechanized in the Rocq prover
using the Iris separation logic framework~\citep{mechanization}.
As future work,
we are interested in three directions.
First, we would like to support less-structured parallelism,
for example programs written using \emph{futures},
and eventually full-blown concurrency.
We posit that the idea of span credits tagged with logical task identifiers
will scale to these settings,
making use of the ``\cantransfername'' approach~(\cref{sec:innertransfer}).
Second, we would like to adapt the expected time credit idea of
\citet{tachis} to allow for proving
bounds on the expected value of work and span in parallel randomized algorithms.
Third, we are also interested in adapting the ideas of \citet{eris}
for proving bounds with high probability on the span.

\begin{acks}
  This work was supported in part by the \grantsponsor{NSF}{National Science Foundation}{} under Grant No.~\grantnum{NSF}{2319168}.
  Any opinions, findings, and conclusions or recommendations expressed in this material are those
  of the authors and do not necessarily reflect the views of the funding agency.
\end{acks}

\bibliography{english, local}

\end{document}